\documentclass[12pt]{article}
\usepackage{epsfig}
\usepackage{amsmath}
\usepackage{amssymb}
\usepackage{amsfonts}
\usepackage{subfigure}

\def\beq{\begin{equation}}
\def\eeq{\end{equation}} 
\def\bea{\begin{eqnarray}}
\def\eea{\end{eqnarray}}
\def\ksl{\hbox{\hbox{${k}$}}\kern-1.9mm{\hbox{${/}$}}}

\newcommand{\nn}{\nonumber}
\def\lsim{\raise0.3ex\hbox{$\;<$\kern-0.75em\raise-1.1ex\hbox{$\sim\;$}}} 

\def\gsim{\raise0.3ex\hbox{$\;>$\kern-0.75em\raise-1.1ex\hbox{$\sim\;$}}}

\begin{document}

\begin{center}

{\bf Solving the Conformal Constraints for Scalar Operators in Momentum Space \\ and the Evaluation of Feynman's Master Integrals}

\vspace{1cm}
{\bf $^a$Claudio Corian\`{o}, $^a$Luigi Delle Rose, $^b$Emil Mottola and $^a$Mirko Serino}

\vspace{1cm}

{\it$^a$ Dipartimento di Matematica e Fisica "Ennio De Giorgi", \\
Universit\`a del Salento\\ and \\
INFN
Lecce, Via Arnesano 73100 Lecce, Italy\\}
\vspace{0.5cm}
{\it$^b$ Theoretical Division, Los Alamos National Laboratory \\Los Alamos, NM, 87545, USA\footnote{claudio.coriano@unisalento.it, luigi.dellerose@le.infn.it, emil@lanl.gov, mirko.serino@le.infn.it}}

\vspace{.5cm}

\begin{abstract} 

We investigate the structure of the constraints on three-point correlation functions emerging when conformal invariance is imposed in 
momentum space and in arbitrary space-time dimensions, presenting a derivation of their solutions for arbitrary scalar operators. 
We show that the differential equations generated by the requirement of symmetry under special conformal transformations coincide with those satisfied by generalized hypergeometric 
functions (Appell's functions). Combined with the position space expression of this correlator, 
whose Fourier transform is given by a family of generalized Feynman (master) integrals, the method allows to derive the expression of such integrals in a completely independent way, 
bypassing the use of Mellin-Barnes techniques, which have been used in the past. 
The application of the special conformal constraints generates a new recursion relation for this family of integrals.  \end{abstract}

\end{center}
\newpage
\section{Introduction} 

Conformal invariance plays an important role in constraining the structure of correlation functions of conformal field theories 
in any dimensions. It allows to fix the form of correlators - up to three-point functions - modulo a set of constants which are 
also given, once the field content of the underlying conformal field theory is selected \cite{Osborn:1993cr, Erdmenger:1996yc}.
The approach, which is largely followed in this case, is naturally formulated in position space, while the same conformal requirements, in 
momentum space, have been far less explored \cite{Coriano:2012wp}.\\
Conformal three-point functions have been intensively studied in the past, and a classification of their possible structures, in the presence of conformal anomalies, is 
available. Conformal anomalies emerge due to the inclusion of the energy momentum tensor in a certain correlator and, in some cases, find specific realizations in free field theories of 
scalars, vectors and fermions \cite{Osborn:1993cr, Erdmenger:1996yc}. Typical correlators which have been studied  are those 
involving the $TT, TOO, TVV$, and $TTT$, where $T$ denotes the energy momentum tensor, $V$ a vector and $O$ a generic 
scalar operator of arbitrary dimension.

The conformal constraints in position space, in this case, are combined with the Ward identities derived from the conservation of the energy momentum tensor and its 
tracelessness condition, valid at separate coordinate points, to fix the structure of each correlator. 
These solutions are obtained for generic conformal theories, with no reference to their Lagrangian realization which, in general, may not even exist. 
The solutions of the conformal constraints are then extended to include the contributions from the coincidence regions, where all the external points collapse to the same 
point. \\
Free field theory realizations of these correlators (for fermions, scalars and vectors) allow to perform a direct test of these results both in position 
and in momentum space, at least in some important cases, such as the $TVV$ or the $TTT$ 
(this latter only for $d=4$) \cite{Coriano:2012wp}, but obviously do not exhaust all possibilities.

Recently interest in the momentum space form of conformal correlators has arisen in the context of the study of anomalous conformal Ward identities, 
massless poles and scalar degrees of freedom associated with the trace anomaly \cite{Giannotti:2008cv, Mottola:2010, Armillis:2009pq, Armillis:2009im, Armillis:2010qk, Coriano:2011ti, Coriano:2011zk}, 
and because of their possible role in determining the form of conformal invariance in the non-Gaussian features of the Cosmic Microwave Background 
\cite{AntMazMott:1997, Antoniadis:2011ib}, or in inflation \cite{Maldacena:2011nz, Kehagias:2012pd}.
The possibility of retrieving information on conformal correlators in momentum space seems to be related, in one way or another, 
to the previous knowledge of the same correlators in configuration space, where the conformal constraints are easier to implement and solve. 
One question that can be naturally raised is if we are able to bypass the study of conformal correlators in position space, by fixing their structure directly in momentum space 
and with no further input. This approach defines an independent path which, as we are going to show, can be successful in some specific cases. 
We will illustrate the direct construction of the solution bringing the example of the scalar three-point correlator. 
The analysis in momentum space should not be viewed though as an unnecessary complication. 
In fact, the solution in the same space, if found, is explicit in the momentum variables and can be immediately compared with the integral representation of the position space solution, 
given by a generalized Feynman integral. As a corollary of this approach, in the case of three-point functions, we are able to determine the complete structure of such an integral, which 
is characterized by three free parameters related to the scaling dimensions of the original scalar operators, in an entirely new way. 
It is therefore obvious that this approach allows to determine the explicit form of an entire family of master integrals.   

In the scalar case we are able to show that the conformal conditions are equivalent to partial differential equations (PDE's) of 
generalized hypergeometric type, solved by functions of two variables, $x$ and $y$, which take the form of ratios of the external momenta.   
The general solution is expressed as a generic linear combination of four generalized hypergeometric functions of the same variables, or Appell's functions. 
Three out of the four constants of the linear combination can be fixed by the momentum symmetry. 
This allows to write down the general form of the scalar correlator in terms of a single multiplicative constant, 
which classifies all the possible conformal realizations of the scalar three-point function. 

In the final part of this work we go back to the analysis of the conformal master integrals, the Fourier transform of the scalar three-point correlators in position space.
We show that the usual rules of integration by parts satisfied by these integrals are nothing else but the requirement of scale invariance. 
Specifically, dilatation symmetry relates the master integral $J(\nu_1,\nu_2,\nu_3)$, labelled by the powers of the Feynman propagators $(\nu_1,\nu_2,\nu_3)$  - with $\nu_1 +\nu_2 +\nu_3=
\kappa$ - to those of the first neighboring plane $(\kappa\to \kappa +1)$.
On the other hand, special conformal constraints relate the integrals of second neighboring planes $(\kappa\to \kappa +2)$. 

%%%%%%%%%%%%%%%%%%%%%%%%%%%%%%%%%%%%%%%%%%%%%%
\section{Conformal transformations}
%%%%%%%%%%%%%%%%%%%%%%%%%%%%%%%%%%%%%%%%%%%%%%
%
In order to render our treatment self-contained, we present a brief review of the conformal transformations in $d > 2$ dimensions which identify, in Minkowski space, the conformal group $SO(2,d)$. \\
These may be defined as the transformations $x_\mu \rightarrow x'_\mu(x)$ that preserve the infinitesimal length up to a local factor 
\bea
d x_\mu d x^\mu \rightarrow d x'_\mu d x'^\mu = \Omega(x)^{-2} d x_\mu d x^\mu \,.
\eea
In the infinitesimal form, for $d > 2$, the conformal transformations are given by
\bea
\label{xtransf}
x'_\mu(x) = x_\mu + a_\mu + {\omega_{\mu}}^{\nu} x_\nu + \lambda x_\mu + b_\mu x^2 - 2 x_\mu b \cdot x
\eea
with
\bea
\label{Omega}
\Omega(x) = 1 -\sigma(x) \qquad \mbox{and} \qquad \sigma(x) = \lambda - 2 b \cdot x \,.
\eea
The transformation in Eq.(\ref{xtransf}) is defined by translations ($a_\mu$), rotations ($\omega_{\mu\nu} = - \omega_{\nu\mu}$), dilatations ($\lambda$) and 
special conformal transformations ($b_\mu$). The first two define the Poincar\'{e} subgroup which leaves invariant the infinitesimal length ($\Omega(x) = 1$). \\
If we also consider the inversion
\bea
x_\mu \rightarrow x'_\mu = \frac{x_\mu}{x^2} \,, \qquad \qquad \Omega(x) = x^2 \,,
\eea 
we can enlarge the conformal group to $O(2,d)$. Special conformal transformations can be realized by a translation preceded and 
followed by an inversion.

Having specified the elements of the conformal group, we can define a quasi primary field $\mathcal O^i(x)$, where the index $i$ runs over a representation of the group $O
(1,d-1)$ to which the field belongs, through the transformation property under a conformal transformation $g$
\bea
\mathcal O^i(x) \stackrel{g}{\rightarrow} \mathcal O'^i(x') = \Omega(x)^\eta D^i_j(g) \mathcal O^j(x) \,,
\eea
where $\eta$ is the scaling dimension of the field and $D^i_j(g)$ denotes the representation of $O(1,d-1)$.
In the infinitesimal form we have
\bea
\delta_g \mathcal O^i(x) = - (L_g \mathcal O)^i(x) \,, \qquad \mbox{with} \qquad  L_g = v \cdot \partial + \eta \, \sigma + \frac{1}{2} \partial_{[ \mu} v_{\nu ]} \Sigma^{\mu\nu} \,,
\eea
where the vector $v_\mu$ is the infinitesimal coordinate variation $v_\mu = \delta_g x_\mu = x'_\mu(x) - x_\mu$ and 
$(\Sigma_{\mu\nu})^i_j$ are the generators of $O(1,d-1)$ in the representation of the field $\mathcal O^i$. The explicit form of the 
operator $L_g$ can be obtained from Eq.(\ref{xtransf}) and Eq.(\ref{Omega}) and is given by
\begin{align}
& \mbox{translations:} && L_g = a^\mu \partial_\mu \,, \nn \\
& \mbox{rotations:} && L_g = \frac{\omega^{\mu\nu}}{2} \left[ x_\nu \partial_\mu - x_\mu \partial_\nu - \Sigma_{\mu\nu} \right] \,, \nn \\
& \mbox{scale transformations :}    && L_g = \lambda \left[ x \cdot \partial +  \eta \right] \,, \nn  \\
& \mbox{special conformal transformations. :}    && L_g = b^\mu \left[ x^2 \partial_\mu - 2 x_\mu \, x \cdot \partial - 2 \eta \, x_\mu - 2 x_\nu {\Sigma_{\mu}}^{\nu} \right] \,.
\end{align}
Conformal invariant correlation functions of quasi primary fields can be defined by requiring that 
\bea
\sum_{r=1}^{n} \langle \mathcal O^{i_1}_1(x_1) \ldots \delta_g \mathcal O^{i_r}_r(x_r) \ldots \mathcal O^{i_n}_n(x_n) \rangle = 0 \,.
\eea
In particular, the invariance under scale and special conformal transformations, in which we are mainly interested, reads as
\bea
\label{ConformalEqCoord}
&& \sum_{r=1}^{n} \left( x_r \cdot \partial^{x_r} +  \eta_r  \right) \langle \mathcal O^{i_1}_1(x_1) \ldots   \mathcal O^{i_r}_r(x_r) 
\ldots  \mathcal O^{i_n}_n(x_n) \rangle = 0 \,, \nn \\
&& \sum_{r=1}^{n} \left( x_r^2 \partial^{x_r}_\mu - 2 x_{r \, \mu} \, x_r \cdot \partial^{x_r} - 2 \eta_r \, x_{r \, \mu} - 2 x_{r \, 
\nu} ( {\Sigma^{(r)}_{\mu}}^{\nu})^{i_r}_{j_r} \right) \langle \mathcal O^{i_1}_1(x_1) \ldots  \mathcal O^{j_r}_r(x_r) \ldots \mathcal 
O^{i_n}_n(x_n) \rangle = 0 \,. \nn \\
\eea
The constraints provided by conformal invariance have been solved in coordinate space and for arbitrary space-time dimension. One can show, 
for instance, that the two and three-point functions are completely determined by conformal symmetry up to a small number of independent
constants \cite{Polyakov:1970, Osborn:1993cr} . \\
Exploiting the same constraints in momentum space is somewhat more involved. 
In the following we assume invariance under the Poincar\'{e} group and we focus our attention on dilatations and special conformal transformations.\\
For this purpose we define the Fourier transform of a $n$ point correlation function as
\bea
&& (2 \pi)^d \, \delta^{(d)}(p_1 + \ldots + p_n) \, \langle \mathcal O^{i_1}_1(p_1) \ldots   \mathcal O^{i_n}_n(p_n) \rangle  \nn \\
&& \qquad \qquad \qquad = \int d^d x_1 \ldots d^d x_n \, \langle \mathcal O^{i_1}_1(x_1) \ldots    \mathcal O^{i_n}_n(x_n) \rangle e^{ 
i p_1 \cdot x_1 + \ldots + i p_n \cdot x_n},
\eea
where the correlation function in momentum space is understood to depend only on $n-1$ momenta, 
as the $n$-th one is removed using momentum conservation. \\
The momentum space differential equations describing the invariance under dilatations and special conformal 
transformations are obtained Fourier-transforming Eq.(\ref{ConformalEqCoord}). It is worth noting 
that some care must be taken, due to the appearance of derivatives on the delta function. As pointed 
out in \cite{Maldacena:2011nz}, these terms can be discarded and we are left with the two equations
\bea
\label{ConformalEqMom}
&& \bigg[ - \sum_{r=1}^{n-1} \left(  p_{r \, \mu} \, \frac{\partial}{\partial p_{r \, \mu}}  + d  \right) + \sum_{r=1}^{n} \eta_r 
\bigg] \langle \mathcal O^{i_1}_1(p_1) \ldots   \mathcal O^{i_r}_r(p_r) \ldots  \mathcal O^{i_n}_n(p_n) \rangle = 0 \,, \nn \\
&& \sum_{r=1}^{n-1} \left( p_{r \, \mu} \, \frac{\partial^2}{\partial p_{r}^{\nu} \partial p_{r \, \nu}}  - 2 \, p_{r \, \nu} \, 
\frac{\partial^2}{ \partial p_{r}^{\mu} \partial p_{r \, \nu} }    + 2 (\eta_r - d) \frac{\partial}{\partial p_{r}^{\mu}}  + 2 
(\Sigma_{\mu\nu}^{(r)})^{i_r}_{j_r} \frac{\partial}{\partial p_{r \, \nu}} \right) \nn \\
&& \hspace{7cm} \, \times \langle \mathcal O^{i_1}_1(p_1) \ldots  \mathcal O^{j_r}_r(p_r) \ldots \mathcal O^{i_n}_n(p_n) \rangle = 0 \, ,
\eea
which define an arbitrary conformal invariant correlation function in $d$ dimensions.
Note that we are dealing with a first and a second order partial differential equations in $n-1$ independent momenta. The choice of the momentum which is eliminated in the two equations given above is arbitrary. \\
Despite the apparent asymmetry in the definition of the special conformal constraint, due to the absence of the $n$-th scaling dimension $\eta_n$ and the $n$-th spin matrix $( \Sigma_{\mu\nu}^{(n)})^{i_n}_{j_n}$, the second of Eq.(\ref{ConformalEqMom}) does not depend on the specific momentum which is eliminated. We could have similarly chosen to express Eq.(\ref{ConformalEqMom}) in terms of the momenta $(p_1 \ldots p_{k-1}, p_{k+1}, \ldots p_n)$, with $p_k$ removed using the momentum conservation, and we would have obtained an equivalent relation. We have left to an appendix the formal proof of this point.
We have then explicitly verified the correctness of our assertion on the vector and tensor two-point correlators which have been discussed in the following section. In both cases, whatever momentum parameterization is chosen for Eq.(\ref{ConformalEqMom}), we have checked that special conformal constraints imply the equality of the scaling dimensions of the two (vector or tensor) operators, accordingly to the well-known result obtained in coordinate space. \\
We recall, anyway, that, 
apart from the following section, in which vector and tensor two-point functions are reviewed, the main results of this work, being focused on scalar operators, are free from all the complications arising from the presence of the spin matrices in the special conformal constraints.

%%%%%%%%%%%%%%%%%%%%%%%%%%%%%%%%%%%%%%%%%%%%%%%%%%%%%%%%%%%%%%%%%%%%%%%%%%%%
\section{Two-point functions from momentum space and anomalies}
\label{TwoPointSection}
%%%%%%%%%%%%%%%%%%%%%%%%%%%%%%%%%%%%%%%%%%%%%%%%%%%%%%%%%%%%%%%%%%%%%%%%%%%%

%%%%%%%%%%%%%%%%%%%%%%%%%%%%%%%%%%%%%%%%%%%%%%%%%%%%%%%%%%%%%%%%%%%%%%%%%%%%
\subsection{General solutions of the scale and special conformal identities}
%%%%%%%%%%%%%%%%%%%%%%%%%%%%%%%%%%%%%%%%%%%%%%%%%%%%%%%%%%%%%%%%%%%%%%%%%%%%

We start exploring the implications of these constraints on two-point functions. 
In particular, the quasi primary fields taken into account are 
scalar ($\mathcal O$), conserved vector ($V_\mu$) and conserved and traceless ($T_{\mu\nu}$) operators.

For the two-point functions the differential equations in Eq.(\ref{ConformalEqMom}) simplify considerably, being expressed in terms 
of just one independent momentum $p$, and take the form
\bea 
\label{ConformalEqMomTwoPoint}
&& \left( - p_{\mu} \, \frac{\partial}{\partial p_{\mu}}  + \eta_1 + \eta_2 - d \right) G^{ij}(p) = 0 \,, \nn \\
&& \left(  p_{\mu} \, \frac{\partial^2}{\partial p^{\nu} \partial p_{\nu}}  - 2 \, p_{\nu} \, \frac{\partial^2}{ \partial p^{\mu} 
\partial p_{\nu} }    + 2 (\eta_1 - d) \frac{\partial}{\partial p^{\mu}}  + 2 (\Sigma_{\mu\nu})^{i}_{k} \frac{\partial}{\partial 
p_{\nu}} \right)  G^{kj}(p)  = 0 \,,
\eea
where we have defined $G^{ij}(p) \equiv \langle \mathcal O_1^i(p) \mathcal O_2^j(-p) \rangle$. The 
first of Eq.(\ref{ConformalEqMomTwoPoint}) dictates the scaling behavior of the correlation function, while special conformal 
invariance allows a non zero result only for equal scale dimensions of the two operators $\eta_1 = \eta_2$, as we know from the 
corresponding analysis in coordinate space. We start by illustrating this point. \\
For the correlation function $G_S(p)$ of two scalar quasi primary fields the invariance under the Poincar\'{e} group obviously 
implies that $G_S(p) \equiv G_S(p^2)$, so that the derivatives with respect to the momentum $p_\mu$ can be easily recast in terms of 
the variable $p^2$. \\
The invariance under scale transformations implies that $G_S(p^2)$ is a homogeneous function of degree 
$\alpha = \frac{1}{2}(\eta_1 + \eta_2 - d)$. 
At the same time, it is easy to show that the second equation in (\ref{ConformalEqMomTwoPoint}) can be satisfied only if $\eta_1 = \eta_2$. 
Therefore conformal symmetry fixes the structure of the scalar two-point function up to an arbitrary overall constant $C$ as
\bea
\label{TwoPointScalar}
G_S(p^2) = \langle \mathcal O_1(p) \mathcal O_2(-p) \rangle = \delta_{\eta_1 \eta_2}  \, C\, (p^2)^{\eta_1 - d/2} \, .
\eea 
If we redefine
\bea
C=c_{S 12} \,  \frac{\pi^{d/2}}{4^{\eta_1 - d/2}} \frac{\Gamma(d/2 - \eta_1)}{\Gamma(\eta_1)} 
\eea
in terms of the new integration constant $c_{S 12}$, the two-point function reads as
\bea 
\label{TwoPointScalar2}
G_S(p^2) =  \delta_{\eta_1 \eta_2}  \, c_{S 12} \,  \frac{\pi^{d/2}}{4^{\eta_1 - d/2}} \frac{\Gamma(d/2 - \eta_1)}{\Gamma(\eta_1)} 
(p^2)^{\eta_1 - d/2} \,,
\eea
and after a Fourier transformation in coordinate space takes the familiar form
\bea
\langle \mathcal O_1(x_1) \mathcal O_2(x_2) \rangle \equiv \mathcal{F.T.}\left[ G_S(p^2) \right] =  \delta_{\eta_1 \eta_2} \,  c_{S 12} 
\frac{1}{(x_{12}^2)^{\eta_1}} \,,
\eea
where $x_{12} = x_1 - x_2$. 
The ratio of the two Gamma functions relating the two integration constants $C$ and $c_{S 12}$ correctly reproduces the ultraviolet singular behavior of the correlation function and plays a role in the discussion of the origin of the scale anomaly.

Now we turn to the vector case where we define $G_V^{\alpha \beta}(p) \equiv \langle V_1^\alpha(p) V_2^\beta(-p) \rangle$. If the 
vector current is conserved, then the tensor structure of the two-point correlation function is entirely fixed by the transversality 
condition, $\partial^\mu V_\mu = 0$, as 
\bea
\label{TwoPointVector0}
G_V^{\alpha \beta}(p) =  \pi^{\alpha\beta}(p) \, f_V(p^2)\,, \qquad \qquad \mbox{with} \qquad
\pi^{\alpha\beta}(p) = \eta^{\alpha \beta} -\frac{p^\alpha p^\beta}{p^2} 
\eea
where $f_V$ is a function of the invariant square $p^2$ whose form, as in the scalar case, is determined by the conformal constraints. 
Following the same reasonings discussed previously we find that
\bea 
\label{TwoPointVector}
G_V^{\alpha \beta}(p) = \delta_{\eta_1 \eta_2}  \, c_{V 12}\, 
\frac{\pi^{d/2}}{4^{\eta_1 - d/2}} \frac{\Gamma(d/2 - \eta_1)}{\Gamma(\eta_1)}\,
\left( \eta^{\alpha \beta} -\frac{p^\alpha p^\beta}{p^2} \right)\
(p^2)^{\eta_1-d/2} \,,
\eea
with $c_{V12}$ being an arbitrary constant. 
We recall that the second equation in (\ref{ConformalEqMomTwoPoint}) gives consistent results for the two-point function in Eq.(\ref{TwoPointVector}) only when the scale dimension $\eta_1 = d - 1$. We refer to appendix \ref{AppTwoPoint} for more details. \\
To complete this short excursus, we present the solution of the conformal constraints for the two-point function built out of two energy momentum tensor operators which are symmetric, conserved and traceless
\bea
\label{EMTconditions}
T_{\mu\nu} = T_{\nu\mu} \,, \qquad \qquad \partial^{\mu} T_{\mu\nu} = 0 \,, \qquad \qquad {T_{\mu}}^{\mu} = 0 \,.
\eea
Exploiting the conditions defined in Eq.(\ref{EMTconditions}) we can unambiguously define the tensor structure of the correlation 
function $G^{\alpha\beta\mu\nu}_T(p) = \Pi_{d}^{\alpha\beta\mu\nu}(p) \, f_T(p^2)$ with
\bea 
\label{TT}
\Pi^{\alpha\beta\mu\nu}_{d}(p) = \frac{1}{2} \bigg[ \pi^{\alpha\mu}(p) \pi^{\beta\nu}(p) + \pi^{\alpha\nu}(p) \pi^{\beta\mu}(p) 
\bigg] 
- \frac{1}{d-1} \pi^{\alpha\beta}(p) \pi^{\mu\nu}(p) \,,
\eea
and the scalar function $f_T(p^2)$ determined as usual, up to a multiplicative constant, by requiring the invariance under 
dilatations and special conformal transformations. We obtain
\bea 
\label{TwoPointEmt}
G^{\alpha\beta\mu\nu}_T(p) = \delta_{\eta_1 \eta_2}  \, 
c_{T 12}\,\frac{\pi^{d/2}}{4^{\eta_1 - d/2}} \frac{\Gamma(d/2 - \eta_1)}{\Gamma(\eta_1)}\, 
\Pi^{\alpha\beta\mu\nu}_{d}(p) \, (p^2)^{\eta_1 - d/2} \,.
\eea
As for the conserved vector currents, also for the energy momentum tensor the scaling dimension is fixed by the second of Eq.(\ref{ConformalEqMomTwoPoint})
and it is given by $\eta_1 = d$. This particular value ensures that $\partial^\mu T_{\mu\nu}$ is also a quasi primary (vector) field. 
We have left to the appendix \ref{AppTwoPoint} the details of the characterization of the vector and tensor two-point functions.\\
These formulae agree with those in the literature \cite{CapFriedLaT:1991}, and in particular those in Sec. 8 of Ref. \cite{Antoniadis:2011ib}
for the gravitational wave spectrum of the CMB.

%%%%%%%%%%%%%%%%%%%%%%%%%%%%%%%%%%%%%%%%%%%%%%%%%%%%%%%%%%%%%%%%%%%%%%%%%%%%%%%%%
\subsection{Divergences and anomalous breaking of scale identities}
%%%%%%%%%%%%%%%%%%%%%%%%%%%%%%%%%%%%%%%%%%%%%%%%%%%%%%%%%%%%%%%%%%%%%%%%%%%%%%%%%

The expressions obtained so far for the two-point functions in Eq.(\ref{TwoPointScalar2}),(\ref{TwoPointVector}) and (\ref{TwoPointEmt}),
allow to discuss very easily the question of the divergences and of the corresponding violations that these induce in the scale identities.
We can naturally see this noting that the Gamma function has simple poles for non positive integer arguments, which occur, in our case, when $\eta = d/2 + n$ with $n=0,1,2,\ldots$. \\
Working in dimensional regularization, we can parametrize the divergence through an analytic continuation of the space-time dimension, $d \to d - 2 \epsilon$, and, then, expand the product $\Gamma(d/2-\eta)\,(p^2)^{\eta - d/2}$, which appears in every two-point function, in a Laurent series around $d/2 - \eta = -n$. We obtain
\bea
\label{expansion}
\Gamma\left(d/2-\eta\right)\,(p^2)^{\eta-d/2} = \frac{(-1)^n}{n!} \left( - \frac{1}{\epsilon} + \psi(n+1)  + O(\epsilon) \right) (p^2)^{n + \epsilon} \,,
\eea
where $\psi(z)$ is the logarithmic derivative of the Gamma function, and $\epsilon$ takes into account the divergence of the two-point correlator for particular values of the scale dimension $\eta$ and of the space-time dimension $d$.

The singular behavior described in Eq.(\ref{expansion}) is responsible for the anomalous 
violation of scale invariance \cite{BrownCol:1980}, providing an extra contribution to the differential equation (\ref{ConformalEqMomTwoPoint}) 
obtained from the conformal symmetry constraints. Indeed, when $\eta = d/2 + n$, employing dimensional regularization, the first of 
Eq.(\ref{ConformalEqMomTwoPoint}) becomes
\bea
\label{AnomScale1}
\left( p^2\, \frac{\partial}{\partial p^2} - n - \epsilon \right)\, G^{ij}(p^2) = 0\, , \qquad \mbox{with} \quad \eta_1 = \eta_2 
\equiv \eta
\eea
which is the Euler equation for a function $G^{ij}$ which behaves like $(p^2)^{n+ \epsilon}$. Due to the appearance of a divergence in $1/\epsilon$
in the correlation function, Eq.(\ref{AnomScale1}) acquires an anomalous finite term 
in the limit $\epsilon \to 0$ and we obtain
\bea
\label{AnomScale2}
\left( p^2\, \frac{\partial}{\partial p^2} - n  \right)\, G^{ij}(p^2) = G^{ij}_{sing}(p^2) \,, 
\eea
where $G^{ij}_{sing}(p^2)$ corresponds to the singular contribution in the correlation function, which we have decomposed according to
\bea
G^{ij}(p^2) = \frac{1}{\epsilon} G^{ij}_{sing}(p^2) + G^{ij}_{finite}(p^2)  \,.
\eea
As one can see from the r.h.s. of Eq.(\ref{AnomScale2}), the coefficient of the divergence, $G^{ij}_{sing}(p^2)$, of the two-point function provides the source for its anomalous scaling. 

We illustrate the points discussed so far with some examples.
Consider, for instance, the scalar correlator in Eq.(\ref{TwoPointScalar2}) with scaling dimension $\eta_1 = \eta_2 \equiv \eta = d/2$. Due to the appearance of a pole in the Gamma function, the two-point correlator develops a divergence and becomes
\bea
\label{2PFscalarDiv}
G_S(p^2) = - \, c_{S12} \, \frac{\pi^{d/2}}{\Gamma \left( d/2 \right)} \left[ \frac{1}{\bar \epsilon} + \log p^2 \right] \,,
\eea
where we have defined for convenience
\bea
\label{epsbar}
\frac{1}{\bar \epsilon} = \frac{1}{\epsilon} + \gamma - \log(4 \pi)  \,,
\eea
with $\gamma$ being the Euler-Mascheroni constant. It is implicitly understood that the argument of the logarithm in Eq.(\ref{2PFscalarDiv}) is made dimensionless, in dimensional regularization, by the insertion of a massive parameter. \\
As one can easily verify, the scalar two-point function given in Eq.(\ref{2PFscalarDiv}) satisfies the anomalous scaling equation (\ref{AnomScale2}) with a constant source term 
\bea
G_{S,sing}(p^2) = - \, c_{S12} \, \frac{\pi^{d/2}}{\Gamma \left( d/2 \right)} 
\eea
determined by the coefficient of the singularity. Note that the anomalous scaling behavior in Eq.(\ref{AnomScale2}) is reproduced by the logarithmic contribution in Eq.(\ref{2PFscalarDiv}). 

Now we turn to the discussion of a correlation function with two vector currents. As already mentioned, the scaling dimension of the conserved vector operator is fixed at the value $\eta = d-1$. In this case the divergences occur at $d = 2n + 2$ with $n= 0,1,\ldots \,$, so that, for $d>2$, the first singularity appears at $d = 4$.
Therefore the vector two-point function for $d=4$ is
\bea 
\label{2PFvectorDiv}
G^{\alpha\beta}_V(p^2) = c_{V12}\,\frac{\pi^{2}}{8}\,
p^2\, \left[ \frac{1}{\bar\epsilon} - 1 + \log p^2 \right] \pi^{\alpha\beta}(p) \,,
\eea
with $\bar \epsilon$ defined in Eq.(\ref{epsbar}).
As for the previous case, it is manifest that the two-point function in Eq.(\ref{2PFvectorDiv}) satisfies the identity given in 
Eq.(\ref{AnomScale2}), with the logarithm accounting for the source of the anomalous scaling behavior. 

Finally, we illustrate the case of the correlation function built with two (symmetric, conserved and traceless) energy momentum 
tensors with scale dimension $\eta=d$,
which is slightly more involved, as we have to pay attention to the fact that $\Pi_d^{\alpha\beta\mu\nu}(p)$ itself depends on the 
space-time dimension $d$. 
The singularities are generated when $d = 2 n$ with $n= 0,1,\ldots \,$, namely for even values of the space-time dimension.
For instance, the two-point function in $d=4$ is found to be given by
\bea
\label{2PFtensorDiv}
G_T^{\alpha\beta\mu\nu}(p) = 
- c_{T12}\, \frac{\pi^2}{192}\, (p^2)^2\,  \bigg\{
\left[ \frac{1}{\bar\epsilon} - \frac{3}{2} + \log p^2  \right]\, \Pi^{\alpha\beta\mu\nu}_4(p)
- \frac{2}{9}\, \pi^{\alpha\beta}(p)\pi^{\mu\nu}(p)
\bigg\} \, .
\eea
As we have already discussed previously, the appearance of the singularity in the correlation function develops an anomalous term in 
the scale identity. Correspondingly, being the energy momentum tensor related to the dilatation current, $J_D^\mu = x_\nu 
T^{\mu\nu}$, it acquires an anomalous trace reflecting the violation of the scale symmetry. In this respect, the two-point function 
in Eq.(\ref{2PFtensorDiv}) is characterized by a non vanishing trace
\bea \label{trace}
\eta_{\mu\nu} \, G_T^{\alpha\beta\mu\nu}(p) =  c_{T12}\, \frac{\pi^2}{288}\, (p^2)^2\, \pi^{\alpha\beta}(p) \,,
\eea 
generated by the last term in Eq.(\ref{2PFtensorDiv}) which, on the other hand, arises from the explicit dependence of the 
$\Pi_d^{\alpha\beta\mu\nu}(p)$ tensor on the space-time dimension. 
The non-zero trace of Eq.(\ref{trace}) is the signature of a conformal or trace anomaly, whose coefficients are known for free fields 
\cite{Duff:1977ay,AndMolMott:2003}.

%%%%%%%%%%%%%%%%%%%%%%%%%%%%%%%%%%%%%%%%%%%%%%%%%%%%%%%%%%%%%%
\section{Three-point functions for scalar operators}
%%%%%%%%%%%%%%%%%%%%%%%%%%%%%%%%%%%%%%%%%%%%%%%%%%%%%%%%%%%%%%

In this section we turn to the momentum space analysis of conformal invariant three-point functions, by solving the constraints emerging 
from the invariance under the conformal group. %
We consider scalar quasi primary fields $\mathcal O_i$ with scale dimensions $\eta_i$ and define the three-point function
\bea
G_{123}(p_1,p_2) = \langle \mathcal O_1(p_1) \mathcal O_2(p_2) \mathcal O_3(-p_1 - p_2) \rangle \,.
\eea
The three-point correlator is a function of the two independent momenta $p_1$ and $p_2$, from which one can construct 
three independent scalar quantities, namely $p_1^2$, 
$p_2^2$ and $p_1 \cdot p_2$. We trade the last invariant for $p_3^2$  in order to manifest the symmetry properties of $G_{123}$ under
the exchange of any couple of operators. \\
We observe that scale invariance, the first equation in Eq.(\ref{ConformalEqMom}), implies that $G_{123}$ is a homogeneous 
function of degree $\alpha = -d + \frac{1}{2}(\eta_1 + \eta_2 + \eta_3)$. Therefore it can be written in the form
\bea
\label{DilatationSol}
G_{123}(p_1^2, p_2^2, p_3^2) = (p_3^2)^{-d + \frac{1}{2}(\eta_1 + \eta_2 + \eta_3)} \, \Phi(x,y)  \qquad \mbox{with} \qquad x = 
\frac{p_1^2}{p_3^2} \,, \quad y = \frac{p_2^2}{p_3^2} \,,
\eea
where we have introduced the dimensionless ratios $x$ and $y$, which must not be confused with coordinate points. The dilatation 
equation only fixes the scaling behavior of the three-point correlator giving no further information on the dimensionless function
$\Phi(x,y)$. 

The last equation of (\ref{ConformalEqMom}), which describes the invariance under special conformal transformations, is the most
predictive one and, as we shall see, completely determines $\Phi(x,y)$ up to a multiplicative constant. \\
To show this, we start by rewriting Eq.(\ref{ConformalEqMom}) in a more useful form by introducing a change of variables from $(p_1^2,p_2^2,p_3^2)$ to 
$(x,y,p_3^2)$. The derivatives respect to the momentum components are re-expressed in terms of derivatives of the momentum invariants and their ratios as
\bea
\frac{\partial}{\partial p_{1}^{\mu}}  &=&   2 (p_{1\, \mu} + p_{2 \, \mu}) \frac{\partial}{\partial p_3^2} + \frac{2}{p_3^2}\left( 
(1- x) p_{1 \, \mu}  - x  \,  p_{2 \, \mu} \right) \frac{\partial}{\partial x} - 2  (p_{1\, \mu} + p_{2 \, \mu}) \frac{y}{p_3^2} 
\frac{\partial}{\partial y} \,, \nn \\
\frac{\partial}{\partial p_{2}^{\mu}}  &=& 2 (p_{1\, \mu} + p_{2 \, \mu}) \frac{\partial}{\partial p_3^2}   -   2  (p_{1\, \mu} + 
p_{2 \, \mu}) \frac{x}{p_3^2} \frac{\partial}{\partial x}   + \frac{2}{p_3^2}\left( (1- y) p_{2 \, \mu}  - y  \,  p_{1 \, \mu} 
\right) \frac{\partial}{\partial y}. \, 
\eea
Similar but lengthier formulas hold for second derivatives. Also notice that the derivatives with respect to $p_3^2$ can 
be removed using the solution of the dilatation constraint in Eq.(\ref{DilatationSol}). Therefore we are left with a differential 
equation in the two dimensionless variables $x$ and $y$. \\
Due to the vector nature of the special conformal transformations, Eq.(\ref{ConformalEqMom}) can be projected out on the two 
independent momenta $p_1$ and $p_2$, obtaining a system of two coupled second order partial differential equations (PDE) for the 
function $\Phi(x,y)$. After several non trivial manipulations, these can be recast in the simple form
\bea
\label{F4diff.eq}
\begin{cases}
 \bigg[ x(1-x) \frac{\partial^2}{\partial x^2} - y^2 \frac{\partial^2}{\partial y^2} - 2 \, x \, y \frac{\partial^2}{\partial x \partial y} +  \left[ \gamma - (\alpha + \beta + 1) x \right] \frac{\partial}{\partial x} \nn \\
\hspace{8cm} - (\alpha + \beta + 1) y \frac{\partial}{\partial y}  - \alpha \, \beta \bigg] \Phi(x,y) = 0 \,, \nn \\
 \bigg[ y(1-y) \frac{\partial^2}{\partial y^2} - x^2 \frac{\partial^2}{\partial x^2} - 2 \, x \, y \frac{\partial^2}{\partial x \partial y} +  \left[ \gamma' - (\alpha + \beta + 1) y \right] \frac{\partial}{\partial y} \nn \\
\hspace{8cm} - (\alpha + \beta + 1) x \frac{\partial}{\partial x}  - \alpha \, \beta \bigg] \Phi(x,y) = 0 \,, 
\end{cases} 
\\
\eea
with the parameters $\alpha, \beta, \gamma, \gamma'$ defined in terms of the scale dimensions of the three scalar operators as
\begin{align}
\label{coeffF4}
& \alpha = \frac{d}{2} - \frac{ \eta_1 + \eta_2 - \eta_ 3 }{2} \,,  && \gamma = \frac{d}{2} - \eta_1 + 1 \,, \nn \\
& \beta = d - \frac{\eta_1 + \eta_2 + \eta_3}{2}  \,, && \gamma' = \frac{d}{2} - \eta_2 + 1 \,.
\end{align}
It is interesting to observe that the system of equations in (\ref{F4diff.eq}), coming from the invariance under special 
conformal transformations, is exactly the system of partial differential equations defining the hypergeometric Appell's function of 
two variables, $F_4(\alpha, \beta; \gamma, \gamma'; x, y)$, with coefficients given in Eq.(\ref{coeffF4}). 
The Appell's function $F_4$ is defined as the double series (see, e.g., \cite{AppellBook,BaileyBook,SlaterBook} for thorough 
discussions of the hypergeometric functions and their properties)
\bea
\label{F4def}
F_4(\alpha, \beta; \gamma, \gamma'; x, y) = \sum_{i = 0}^{\infty}\sum_{j = 0}^{\infty} \frac{(\alpha)_{i+j} \, 
(\beta)_{i+j}}{(\gamma)_i \, (\gamma')_j} \frac{x^i}{i!} \frac{y^j}{j!} 
\eea
where $(\alpha)_i = \Gamma(\alpha + i)/ \Gamma(\alpha)$ is the Pochhammer symbol. \\ 
It is known that the system of partial differential equations (\ref{F4diff.eq}), besides the hypergeometric function introduced in 
Eq.(\ref{F4def}), has three other independent solutions given by
\bea
\label{solutions}
S_2(\alpha, \beta; \gamma, \gamma'; x, y) &=& x^{1-\gamma} \, F_4(\alpha-\gamma+1, \beta-\gamma+1; 2-\gamma, \gamma'; x,y) \,, \nn \\
S_3(\alpha, \beta; \gamma, \gamma'; x, y) &=& y^{1-\gamma'} \, F_4(\alpha-\gamma'+1,\beta-\gamma'+1;\gamma,2-\gamma' ; x,y) \,, \nn \\
S_4(\alpha, \beta; \gamma, \gamma'; x, y) &=& x^{1-\gamma} \, y^{1-\gamma'} \, 
F_4(\alpha-\gamma-\gamma'+2,\beta-\gamma-\gamma'+2;2-\gamma,2-\gamma' ; x,y) \, . \nn \\
\eea
Therefore the function $\Phi(x,y)$, solution of (\ref{F4diff.eq}), is a linear combination of the four independent hypergeometric 
functions, i.e.
\bea
\label{SCTSol}
G_{123}(p_1^2, p_2^2, p_3^2) &=& (p_3^2)^{-d + \frac{1}{2}(\eta_1 + \eta_2 + \eta_3)} \, \Phi(x,y) \nn \\
&=& (p_3^2)^{-d + \frac{1}{2}(\eta_1 + \eta_2 + \eta_3)} \sum_{i=1}^{4} c_i(\eta_1,\eta_2,\eta_3) \, S_i(\alpha, \beta; \gamma, 
\gamma'; x, y) \,,
\eea
where we have denoted with $S_1$ the Appell's function $F_4$ given in Eq.(\ref{F4def}), while the parameters 
$\alpha,\beta,\gamma,\gamma'$ are defined in Eq.(\ref{coeffF4}). The $c_i(\eta_1,\eta_2,\eta_3)$ appearing in the linear combination, 
are the arbitrary coefficients which may depend on the scale dimensions $\eta_i$ of the quasi primary fields and on the space-time 
dimension $d$. \\
The coefficients $c_i(\eta_1,\eta_2,\eta_3)$ can be determined, up to an overall multiplicative constant, by exploiting the symmetry 
of the correlation function under the interchange of two of the three scalar operators present in the correlator, which consists of the simultaneous exchange of momenta and 
scale dimensions $(p_i^2, \eta_i) \leftrightarrow (p_j^2,\eta_j)$. \\
Consider, for instance, the invariance of the three-point function under the exchange $\mathcal O_2(p_2) \leftrightarrow \mathcal 
O_3(-p_1-p_2)$, which is achieved by $(p_2^2, \eta_2) \leftrightarrow (p_3^2,\eta_3)$. Then Eq.(\ref{SCTSol}) becomes
\bea
G_{132}(p_1^2, p_3^2, p_2^2) = (p_2^2)^{-d + \frac{1}{2}(\eta_1 + \eta_2 + \eta_3)} \sum_{i=1}^{4} c_i(\eta_1,\eta_3,\eta_2) \, S_i(\tilde \alpha, \tilde \beta; \tilde \gamma, \tilde \gamma'; \frac{x}{y}, \frac{1}{y}) \,,
\eea
where
\begin{align}
& \tilde \alpha = \alpha(\eta_2 \leftrightarrow \eta_3) = \frac{d}{2} - \frac{ \eta_1 + \eta_3 - \eta_ 2 }{2} \,,  && \tilde \gamma = \gamma(\eta_2 \leftrightarrow \eta_3) = \frac{d}{2} - \eta_1 + 1 = \gamma \,, \nn \\
& \tilde \beta = \beta(\eta_2 \leftrightarrow \eta_3) = d - \frac{\eta_1 + \eta_2 + \eta_3}{2} = \beta \,, && \tilde \gamma' = \gamma'(\eta_2 \leftrightarrow \eta_3) = \frac{d}{2} - \eta_3 + 1 \,.
\end{align}
Note that the hypergeometric functions are now evaluated in $x/y$ and $1/y$. To reintroduce the dependence from $x$ and $y$, in order to exploit more easily the symmetry relation
\bea
G_{123}(p_1^2, p_2^2, p_3^2) = G_{132}(p_1^2, p_3^2, p_2^2) \,,
\eea 
we make use of the transformation property of $F_4$ \cite{AppellBook}
\bea
\label{transfF4}
F_4(\alpha, \beta; \gamma, \gamma'; x, y) &=& \frac{\Gamma(\gamma') \Gamma(\beta - \alpha)}{ \Gamma(\gamma' - \alpha) \Gamma(\beta)} (- y)^{- \alpha} \, F_4(\alpha, \alpha -\gamma' +1; \gamma, \alpha-\beta +1; \frac{x}{y}, \frac{1}{y}) \nn \\
&+&  \frac{\Gamma(\gamma') \Gamma(\alpha - \beta)}{ \Gamma(\gamma' - \beta) \Gamma(\alpha)} (- y)^{- \beta} \, F_4(\beta -\gamma' +1, \beta ; \gamma, \beta-\alpha +1; \frac{x}{y}, \frac{1}{y}) \,.
\eea
After some algebraic manipulations, and repeating the procedure described so far for the other operator interchanges, the ratios between the coefficients $c_i$ take the simplified form
\bea
\frac{c_1(\eta_1, \eta_2, \eta_3)}{c_3(\eta_1, \eta_2, \eta_3)} &=& 
\frac{\Gamma \left(\eta _2-\frac{d}{2}\right) \Gamma
   \left(d-\frac{\eta _1}{2}-\frac{\eta _2}{2}-\frac{\eta
   _3}{2}\right) \Gamma \left(\frac{d}{2}-\frac{\eta
   _1}{2}-\frac{\eta _2}{2}+\frac{\eta _3}{2}\right)}{\Gamma \left(\frac{d}{2}-\eta _2\right) \Gamma
   \left(-\frac{\eta _1}{2}+\frac{\eta _2}{2}+\frac{\eta
   _3}{2}\right)  \Gamma
   \left(\frac{d}{2}-\frac{\eta _1}{2}+\frac{\eta
   _2}{2}-\frac{\eta _3}{2}\right)} \,, \nn \\
\frac{c_2(\eta_1, \eta_2, \eta_3)}{c_4(\eta_1, \eta_2, \eta_3)} &=& 
\frac{\Gamma \left(\eta
   _2-\frac{d}{2}\right)  \Gamma \left(\frac{\eta _1}{2}-\frac{\eta
   _2}{2}+\frac{\eta _3}{2}\right) \Gamma \left(\frac{d}{2}+\frac{\eta
   _1}{2}-\frac{\eta _2}{2}-\frac{\eta _3}{2}\right)}{ \Gamma \left(\frac{d}{2}-\eta _2\right) \Gamma
   \left(\frac{\eta _1}{2}+\frac{\eta _2}{2}-\frac{\eta
   _3}{2}\right) \Gamma
   \left(-\frac{d}{2}+\frac{\eta _1}{2}+\frac{\eta
   _2}{2}+\frac{\eta _3}{2}\right)} \,, \nn \\
\frac{c_1(\eta_1, \eta_2, \eta_3)}{c_4(\eta_1, \eta_2, \eta_3)} &=& 
\frac{\Gamma \left(\eta _1-\frac{d}{2}\right) \Gamma \left(\eta
   _2-\frac{d}{2}\right) \Gamma \left(d-\frac{\eta
   _1}{2}-\frac{\eta _2}{2}-\frac{\eta _3}{2}\right) \Gamma
   \left(\frac{d}{2}-\frac{\eta _1}{2}-\frac{\eta
   _2}{2}+\frac{\eta _3}{2}\right)}{\Gamma
   \left(\frac{d}{2}-\eta _1\right) \Gamma
   \left(\frac{d}{2}-\eta _2\right) \Gamma \left(\frac{\eta
   _1}{2}+\frac{\eta _2}{2}-\frac{\eta _3}{2}\right) \Gamma
   \left(-\frac{d}{2}+\frac{\eta _1}{2}+\frac{\eta
   _2}{2}+\frac{\eta _3}{2}\right)} \,,
\eea
and define $G_{123}(p_1^2, p_2^2, p_3^2)$ up to a multiplicative arbitrary constant $c_{123} \equiv c_{123}(\eta_1, \eta_2, \eta_3)$. This depends on the space-time dimension $d$, on the scale dimensions $\eta_i$ of the quasi primary fields and on their normalization. \\
The conformal invariant correlation function of three scalar quasi primary fields with arbitrary scale dimensions is then given by
\small
\bea
\label{ThreePointScalar}
G_{123}(p_1^2, p_2^2, p_3^2) &=&  \frac{c_{123} \,\, \pi^d \, 4^{d - \frac{1}{2}(\eta_1 + \eta_2 + \eta_3)} \, (p_3^2)^{-d + \frac{1}{2}(\eta_1 + \eta_2 + \eta_3)}}{\Gamma\left( \frac{\eta_1}{2} + \frac{\eta_2}{2} - \frac{\eta_3}{2}\right) \Gamma\left( \frac{\eta_1}{2} - \frac{\eta_2}{2} + \frac{\eta_3}{2}\right) \Gamma\left( - \frac{\eta_1}{2} + \frac{\eta_2}{2} + \frac{\eta_3}{2}\right) \Gamma\left( - \frac{d}{2} + \frac{\eta_1}{2} + \frac{\eta_2}{2} + \frac{\eta_3}{2}\right)} \bigg\{ \nn \\
&& \hspace{-2cm} \Gamma \left(\eta _1-\frac{d}{2}\right) \Gamma \left(\eta
   _2-\frac{d}{2}\right) \Gamma \left(d-\frac{\eta
   _1}{2}-\frac{\eta _2}{2}-\frac{\eta _3}{2}\right) \Gamma
   \left(\frac{d}{2}-\frac{\eta _1}{2}-\frac{\eta
   _2}{2}+\frac{\eta _3}{2}\right) \nn \\
&& \hspace{-2cm} \times \, F_4 \left( \frac{d}{2} - \frac{\eta_1 + \eta_2 - \eta_3}{2}, d - \frac{\eta_1 + \eta_2 + \eta_3}{2}; \frac{d}{2} - \eta_1 +1, \frac{d}{2} - \eta_2 +1; x, y \right) \nn \\
&& \hspace{-2cm} +\,   \Gamma \left(\frac{d}{2}-\eta _1\right) \Gamma
   \left(\eta _2-\frac{d}{2}\right) \Gamma \left(\frac{\eta _1}{2}-\frac{\eta _2}{2}+\frac{\eta
   _3}{2}\right) \Gamma
   \left(\frac{d}{2}+\frac{\eta _1}{2}-\frac{\eta
   _2}{2}-\frac{\eta _3}{2}\right) \nn \\
&& \hspace{-2cm} \times \, x^{\eta_1 - \frac{d}{2}} \, F_4\left( \frac{d}{2} - \frac{\eta_2 + \eta_3 - \eta_1}{2}, \frac{\eta_1 + \eta_3 - \eta_2}{2}; - \frac{d}{2} + \eta_1 +1, \frac{d}{2} - \eta_2 +1 ; x, y\right) \nn \\
&& \hspace{-2cm} + \, \Gamma \left(\eta _1-\frac{d}{2}\right) \Gamma
   \left(\frac{d}{2}-\eta _2\right) \Gamma \left(-\frac{\eta _1}{2}+\frac{\eta _2}{2}+\frac{\eta
   _3}{2}\right) \Gamma
   \left(\frac{d}{2}-\frac{\eta _1}{2}+\frac{\eta
   _2}{2}-\frac{\eta _3}{2}\right) \nn \\
&& \hspace{-2cm} \times \, y^{\eta_2 - \frac{d}{2}} \, F_4\left( \frac{d}{2} - \frac{\eta_1 + \eta_3 - \eta_2}{2} , \frac{\eta_2 + \eta_3 - \eta_1}{2}; 
\frac{d}{2} - \eta_1 +1, -\frac{d}{2} + \eta_2 +1 ; x, y\right) \nn \\
&& \hspace{-2cm} + \, \Gamma \left(\frac{d}{2}-\eta _1\right) \Gamma
   \left(\frac{d}{2}-\eta _2\right) \Gamma \left(\frac{\eta _1}{2}+\frac{\eta _2}{2}-\frac{\eta
   _3}{2}\right) \Gamma
   \left(-\frac{d}{2}+\frac{\eta _1}{2}+\frac{\eta
   _2}{2}+\frac{\eta _3}{2}\right) \nn \\
&& \hspace{-2cm} \times \,   x^{\eta_1 - \frac{d}{2}}  y^{\eta_2 - \frac{d}{2}} \, F_4 \left( -\frac{d}{2} + \frac{\eta_1+\eta_2+\eta_3}{2}, 
\frac{\eta_1+\eta_2-\eta_3}{2}; -\frac{d}{2} +\eta_1 +1, -\frac{d}{2} + \eta_2 +1; x, y \right) \bigg\} \,. \nn \\
\eea 
\normalsize
The convenient normalization employed in Eq.(\ref{ThreePointScalar}) for the three-point function reproduces, through the operator 
product expansion, as we are going to show next, the normalization of the two-point functions which we have chosen in 
Eq.(\ref{TwoPointScalar2}). 

As we shall identify the three-point correlator discussed in this section with specific Feynman amplitudes, this will fix the arbitrary constant $c_{123}$ using some information coming from the same operator product expansion analysis. This topic will be presented in section \ref{Section.Davy}. Indeed, the solution of the momentum space version of the conformal constraints provides an alternative computational tool for correlation functions with conformal symmetry.

It is worth to emphasize the connection between the invariance under special conformal transformations and appearance of the Appell's functions. Indeed we have shown how the constraints provided by the conformal group translate, in momentum space, in the well-known system of partial differential equations defining the hypergeometric series $F_4$. We have analyzed this connection in the case of a conformally invariant three-point function built with scalar operators in some detail. A similar correspondence should also hold for more complicated vector and tensor correlators.

%%%%%%%%%%%%%%%%%%%%%%%%%%%%%%%%%%%%%%%%%%%%%%%%%%%%%%%%%%%%%%%
\subsection{The Operator Product Expansion analysis}
%%%%%%%%%%%%%%%%%%%%%%%%%%%%%%%%%%%%%%%%%%%%%%%%%%%%%%%%%%%%%%% 

In this section we show the consistency of our result with the operator product expansion (OPE) in conformal field theories in which
the structure of the Wilson's coefficients is entirely fixed by the scaling dimensions of the two operators. \\
Considering, for instance, the coincidence limit in the scalar case, one has
\bea
\mathcal O_i(x_1) \mathcal O_j(x_2) \sim \sum_k \frac{c_{ijk}}{(x_{12}^2)^{\frac{1}{2}(\eta_i + \eta_j - \eta_k)}} \mathcal O_k(x_2)  
\qquad \mbox{for} \quad x_1 \rightarrow x_2 \,,
\eea
where $x_{12} = x_1 - x_2$. It is worth noting that the coefficients $c_{ijk}$ are the same structure constants 
appearing in the three-point functions. \\
For the correlation function of three scalar operators the OPE implies the singular behavior
\bea
\label{OPEcoord}
\langle \mathcal O_1(x_1) \mathcal O_2(x_2) \mathcal O_3(x_3) \rangle \stackrel{x_3 \rightarrow x_2}{\sim} 
\frac{c_{123}}{(x_{23}^2)^{\frac{1}{2}(\eta_2 + \eta_3- \eta_1)}} \langle \mathcal O_1(x_1) \mathcal O_2(x_2) \rangle \,,
\eea
with analogous formulae for the other coincidence limits. For the sake of simplicity, we choose a diagonal basis of quasi primary operators normalized as
\bea
\langle \mathcal O_{i}(x_1) \mathcal O_{j}(x_2) \rangle = \frac{\delta_{i j}}{(x_{12}^2)^{\eta_i}} \,. 
\eea
The momentum space version of the OPE in Eq.(\ref{OPEcoord}) reads
\bea
\label{OPEmom}
\langle \mathcal O_1(p_1) \mathcal O_2(p_2) \mathcal O_3(-p_1-p_2) \rangle \nn \\
&& \hspace{-4cm } \sim 
\frac{\pi^{d/2}}{4^{\frac{1}{2}(\eta_2+\eta_3-\eta_1)-\frac{d}{2}}} 
\frac{\Gamma(\frac{d}{2}-\frac{\eta_2+\eta_3-\eta_1}{2})}{\Gamma(\frac{\eta_2+\eta_3-\eta_1}{2})} 
\frac{c_{123}}{(p_3^2)^{\frac{d}{2}-\frac{1}{2}(\eta_2+\eta_3-\eta_1)}} \langle \mathcal O_1(p_1) \mathcal O_2(-p_1)\rangle\, ,
\eea
where the scalar two-point function is normalized as in Eq.(\ref{TwoPointScalar2}) with $c_{S12} = 1$. In the previous equation the 
symbol $\sim$ stands for the momentum space counterpart of the short distance limit $x_3 \rightarrow x_2$ which is achieved by the 
$p_3^2, p_2^2 \rightarrow \infty$ limit with $p_2^2 / p_3^2 \rightarrow 1$.  \\
The result for the scalar three-point function given in Eq.(\ref{ThreePointScalar}) is indeed in agreement, as expected, with the OPE 
analysis. This can be shown from Eq.(\ref{ThreePointScalar}) by a suitable expansion of the corresponding Appell's functions. In 
particular, in order to reproduce the momentum space singular behavior of Eq.(\ref{OPEmom}), we need the hypergeometric leading 
expansion in the limit $x = p_1^2/p_3^2 \rightarrow 0$ and $y = p_2^2/p_3^2 \rightarrow 1$, which reads as \cite{AppellBook}
\bea
F_4(\alpha, \beta; \gamma, \gamma'; x, y) \sim \frac{\Gamma(\gamma') \Gamma(\gamma' -\alpha -\beta)}{\Gamma(\gamma' - \alpha) \Gamma(\gamma' - \beta)} \qquad \mbox{for} \quad x \rightarrow 0 \,, y \rightarrow 1 	\,.
\eea
In the previous equation we have retained only the terms with the correct power-law scaling in the $p_3^2$ variable, as dictated by the OPE analysis. In this case these contributions come from the terms of Eq.(\ref{ThreePointScalar}) which are proportional to the $S_2$ and $S_4$ solutions defined in Eq.(\ref{solutions}).
Analogously, in the limit $p_3^2, p_1^2 \rightarrow \infty$, with $p_1^2 / p_3^2 \rightarrow 1$, which is described in coordinate space by $x_3 \rightarrow x_1$, the leading behavior is extracted from $S_3$ and $S_4$. \\
The remaining coincidence limit $x_1 \rightarrow x_2$, corresponding to $p_1^2, p_2^2 \rightarrow \infty$ with $p_1^2 / p_2^2 \rightarrow 1$, is more subtle due to the apparent asymmetry in the momentum invariants $p_1^2, p_2^2, p_3^2$ of the three-point scalar correlator, as given in Eq.(\ref{ThreePointScalar}). In this case both $x$ and $y$ grow to infinity while their ratio $x/y \rightarrow 1$. Therefore it is necessary to apply the transformation defined in Eq.(\ref{transfF4}) to each hypergeometric function appearing in Eq.(\ref{ThreePointScalar}). This can be viewed as an analytic continuation outside the domain of convergence $|\sqrt{x}| + |\sqrt{y}| < 1$, where the Appell's function is strictly defined as a double series.
The hypergeometric functions are then expanded according to
\bea
F_4(\alpha, \beta; \gamma, \gamma'; x, y) & \sim& (- y)^{- \alpha} \frac{ \Gamma(\gamma) \Gamma(\gamma') \Gamma(\beta - \alpha) 
\Gamma(\gamma  + \gamma' - 2 \alpha - 1)}{\Gamma(\beta)  \Gamma(\gamma -\alpha)  
\Gamma(\gamma' - \alpha)\Gamma(\gamma + \gamma' - \alpha - 1)}   \nn \\
&+& 
(- y)^{- \beta}   \frac{\Gamma(\gamma) \Gamma(\gamma') \Gamma(\alpha - \beta)  \Gamma(\gamma + \gamma' - 2 \beta - 1)  }{ 
\Gamma(\alpha)  \Gamma(\gamma' - \beta) \Gamma(\gamma - \beta)  \Gamma(\gamma + \gamma' - \beta - 1) }\, ,  \nn \\
&\mbox{for}& 
 \quad x, y \rightarrow \infty \,, \frac{x}{y} \rightarrow 1 \,. 
\eea
This completes the analysis of the OPE on the three-point scalar function in the three different coincidence limits.

%%%%%%%%%%%%%%%%%%%%%%%%%%%%%%%%%%%%%%%%%%%%%%%%
\section{Feynman integral representation of the momentum space solution}
\label{Section.Davy}
%%%%%%%%%%%%%%%%%%%%%%%%%%%%%%%%%%%%%%%%%%%%%%%%

We have seen in the previous sections that we can fix the explicit structure of the generic three-point scalar correlator 
in momentum space by solving the conformal constraints, which are mapped to a system of two hypergeometric differential equations of two variables. These variables take the form of 
two ratios of the external momenta. 
In particular we find that in any $d$ dimensional conformal field theory the solution of this system of PDE's is characterized by a single integration constant which depends on the 
specific conformal realization, as expected. 

In this section we want to point out the relationship between the scalar three-point functions studied so far and a certain class of Feynman master integrals.
These can be obtained by a Fourier transformation of the corresponding solution of the conformal constraints in coordinate space, which is well known to be
\bea \label{OOO}
\langle \mathcal O_1(x_1) \mathcal O_2(x_2) \mathcal O_3(x_3)\rangle = \frac{c_{123}}{ \left(x_{12}^2\right)^{\frac{1}{2}(\eta_1 +\eta_2-\eta_3)}
\left(x_{23}^2\right)^{\frac{1}{2}(\eta_2 +\eta_3-\eta_1)} \left(x_{31}^2\right)^{\frac{1}{2}(\eta_3 +\eta_1-\eta_2)}}\, .
\eea
Transforming to momentum space, we find an integral representation, which necessarily has to coincide, up to an unconstrained overall constant, 
with the explicit solution found in the previous section, and reads as
\bea
\label{davy}
J(\nu_1,\nu_2,\nu_3) = \int \frac{d^d l}{(2 \pi)^d} \frac{1}{(l^2)^{\nu_3} ((l+p_1)^2)^{\nu_2} ((l-p_2)^2)^{\nu_1}}\, ,
\eea
with external momenta $p_1$, $p_2$ and $p_3$ constrained by momentum conservation $p_1 + p_2 + p_3 = 0$ and
the scale dimensions $\eta_i$ related to the indices $\nu_i$ as
\bea
\label{etafromnu}
\eta_1 = d - \nu_2 - \nu_3 \,, \qquad
\eta_2 = d - \nu_1 - \nu_3 \,, \qquad
\eta_3 = d - \nu_1 - \nu_2 \,. 
\eea
This expression describes a family of master integrals which has been studied in \cite{Boos:1987bg, Davydychev:1992xr},
whose explicit relation with Eq.(\ref{OOO}) is given by
\bea
&& \int \frac{d^d p_1}{(2\pi)^d} \frac{d^d p_2}{(2\pi)^d} \frac{d^d p_3}{(2\pi)^d} \, (2\pi)^d \delta^{(d)}(p_1 + p_2 + p_3) \, 
J(\nu_1,\nu_2,\nu_3) e^{- i p_1 \cdot x_1 - i p_2 \cdot x_2 - i p_3 \cdot x_3} \nn \\
&& = \frac{1}{4^{\nu_1+\nu_2+\nu_3} \pi^{3 d/2}}  \frac{\Gamma(d/2 - \nu_1) \Gamma(d/2 - \nu_2) \Gamma(d/2 - \nu_3)}{\Gamma(\nu_1) 
\Gamma(\nu_2) \Gamma(\nu_3)}  \frac{1}{(x_{12}^2)^{d/2- \nu_3} (x_{23}^2)^{d/2- \nu_1} (x_{31}^2)^{d/2- \nu_2}}\,, \nn \\
\eea
The integral in Eq.(\ref{davy}) satisfies the system of PDE's (\ref{F4diff.eq}). 
Therefore, it can be expressed in terms of the general solution given in Eq.(\ref{ThreePointScalar}) which involves a linear 
combination of four Appell's functions, with the relative coefficients fixed by the symmetry conditions on the dependence from the external momenta.
Then Eq.(\ref{ThreePointScalar}) identifies $J(\nu_1,\nu_2,\nu_3)$ except for an overall  constant $c_{123}$ which we are now going to determine. This task can be 
accomplished, for instance, by exploiting some boundary conditions. \\
As for the OPE analysis discussed in the previous section, we may consider the large momentum limit in which the three-point integral 
collapses into a two-point function topology. 
Taking, for instance, the $p_2^2, p_3^2 \rightarrow \infty$ limit with $p_2^2/p_3^2 \rightarrow 1$ we have
\bea
\label{davylimit}
J(\nu_1,\nu_2,\nu_3) \sim \frac{1}{(p_2^2)^{\nu_1}} \int \frac{d^d l}{(2 \pi)^d} \frac{1}{(l^2)^{\nu_3} ((l + p_1)^2)^{\nu_2}} =  \frac{1}{(p_2^2)^{\nu_1}}  \frac{i^{1-d}}{(4 \pi)^{d/2}} \, G(\nu_2,\nu_3) \, (p_1^2)^{d/2 - \nu_2 - \nu_3} \,,
\eea
where
\bea
G(\nu, \nu') = \frac{\Gamma(d/2 - \nu) \Gamma(d/2 - \nu') \Gamma(\nu+\nu'-d/2)}{\Gamma(\nu) \Gamma(\nu') \Gamma(d - \nu -\nu')} \,.
\eea
Eq.(\ref{davylimit}) must be compared with the same limit taken on the explicit solution in Eq.(\ref{ThreePointScalar}), where the scale dimensions $\eta_i$ are replaced by $\nu_i$ through Eq.(\ref{etafromnu}). This completely determines the multiplicative constant $c_{123}$ and the correct normalization of the three-point master integral, which is obtained by choosing 
\bea
\label{c123}
c_{123} = \frac{i^{1-d}}{4^{\nu_1+\nu_2+\nu_3} \pi^{3 d/2}}  \frac{\Gamma(d/2 - \nu_1) \Gamma(d/2 - \nu_2) \Gamma(d/2 - \nu_3)}{\Gamma(\nu_1) \Gamma(\nu_2) \Gamma(\nu_3)}  \,.
\eea
Therefore the scalar master integral is given by
\bea
J(\nu_1,\nu_2,\nu_3) = G_{123}(p_1^2, p_2^2, p_3^2)
\eea
with scaling dimensions defined in Eq.(\ref{etafromnu}) and the coefficient $c_{123}$ in Eq.(\ref{c123}). Notice that this method allows us to bypass completely the Mellin-Barnes techniques which has been used previously in the analysis of the same integral.

%%%%%%%%%%%%%%%%%%%%%%%%%%%%%%%%%%%%%%%%%%%%%%%
\subsection{Recurrence relations from conformal invariance}
%%%%%%%%%%%%%%%%%%%%%%%%%%%%%%%%%%%%%%%%%%%%%%%

Having established the conformal invariance of the generalized three-point master integral $J(\nu_1,\nu_2,\nu_3)$, we can study the 
implications of the conformal constraints on the integral representation of Eq.(\ref{davy}). These are automatically satisfied by the explicit solution given in Eq.(\ref{ThreePointScalar}), but once that they are applied on 
$J(\nu_1,\nu_2,\nu_3)$, generate recursion relations among the indices of this family of integrals. Specifically, they relate integrals with $\nu_1+\nu_2+\nu_3=\kappa$ to those with $\nu_1+\nu_2+\nu_3=\kappa +1$ and 
$\nu_1+\nu_2+\nu_3=\kappa +2$. For instance, differentiating Eq.(\ref{davy}) under the integration sign according to the first of Eq.(\ref{ConformalEqMom}), which is the condition of scale invariance, 
we easily obtain the recursion relation
\bea 
\label{davyscale}
\nu_2\,p_1^2\,J(\nu_1,\nu_2+1,\nu_3) + \nu_1\,p_2^2\,J(\nu_1+1,\nu_2,\nu_3) 
&=&
\left(\nu_1 + \nu_2 + 2\, \nu_3 - d \right) \, J(\nu_1,\nu_2,\nu_3) \nn \\
&& \hspace{-4cm} + \,
\nu_2\, J(\nu_1,\nu_2+1,\nu_3-1) + \nu_1\, J(\nu_1+1,\nu_2,\nu_3-1) \, , 
\eea
together with the corresponding symmetric relations obtained interchanging 
$(p_1^2, \nu_1) \leftrightarrow (p_3^2, \nu_3)$ or $(p_2^2, \nu_2) \leftrightarrow (p_3^2, \nu_3)$. 
These equations link scalar integrals on two contiguous planes, as mentioned above. 
The recurrence relations obtained from scale invariance exactly correspond to those presented in \cite{Davydychev:1992xr}
and following from the usual integration-by-parts technique, which in this case is derived from the divergence theorem in dimensional regularization
\bea
\int \frac{d^d l}{(2\pi)^d}\, \frac{\partial}{\partial l_{\mu}}\,
\left\{ \frac{l_{\mu}}{(l^2)^{\nu_3}((l+p_1)^2)^{\nu_2}((l-p_2)^2)^{\nu_1}} \right\}= 0.
\label{parts}
\eea
We can easily show the equivalence between Eq.(\ref{parts}) and the first of Eq.(\ref{ConformalEqMom}) which is the constraint of scale invariance.  
In fact, the scale transformation acts on $J(\nu_1,\nu_2,\nu_3)$ in the form 
\bea \label{davyder}
\left[d - 2\,\left(\nu_1 + \nu_2 + \nu_3 \right)
- p_1\cdot \frac{\partial}{\partial p_1} - p_2\cdot \frac{\partial}{\partial p_2} \right]
\int d^d l   \frac{1}{(l^2)^{\nu_3}\,((l+p_1)^{2})^{\nu_2}\,((l-p_2)^{2})^{\nu_1}} = 0 \, .
\eea
Now we just invoke Euler's theorem on homogeneous functions on the integrand, which is of degree 
$- 2\,(\nu_1+\nu_2+\nu_3)$ in the momenta $p_1$, $p_2$ and $l$  and obtain the relation
\bea
&&
\left[ p_1\cdot \frac{\partial}{\partial p_1} + p_2\cdot \frac{\partial}{\partial p_2}
+ l\cdot \frac{\partial}{\partial l} \right] \frac{1}{(l^2)^{\nu_3}\,((l+p_1)^{2})^{\nu_2}\,((l-p_2)^{2})^{\nu_1}} \nn \\\
&& \hspace{8cm}
= \frac{-2 (\nu_1+\nu_2+\nu_3) }{(l^2)^{\nu_3}\,((l+p_1)^{2})^{\nu_2}\,((l-p_2)^{2})^{\nu_1}} \, .
\label{euler}
\eea
At this point, if we combine Eqs.(\ref{davyder}) and (\ref{euler}) and rewrite $d$ as $\frac{\partial}{\partial l} \cdot l$, 
we easily obtain the equivalence with Eq.(\ref{parts}).

Other recursive relations can be found requiring Eq.(\ref{davy}) to satisfy the constraint of special conformal invariance 
which, from the second equation in Eq.(\ref{ConformalEqMom}), takes the form 
\beq
\left\{ p_{1\,\mu} \frac{\partial^2}{\partial p_1 \cdot \partial p_1} - 2\, p_{1\,\nu} \frac{\partial^2}{\partial p_{1}^{\mu}\partial p_{1\,\nu}}
- 2\,(\nu_2+\nu_3)\frac{\partial}{\partial p_{1}^{\mu}}+ (1 \leftrightarrow 2) \right\}\, J(\nu_1,\nu_2,\nu_3) = 0\, .
\label{special}
\eeq
This is a vector condition which involves some tensor integrals of the same $J(\nu_1,\nu_2,\nu_3)$ family.  
Differentiating the integral $J(\nu_1,\nu_2,\nu_3)$ as in Eq.(\ref{special}) and performing some standard manipulations one arrives at the implicit formula
\small
\bea
\label{vectorid}
&& \hspace{-0.6cm}
\nu_2 \, p_{1\, \mu} \bigg[
\left(1 +\nu_2 + \nu_3-d/2\right) J(\nu_1,\nu_2+1,\nu_3)
+ (\nu_2+1) \left( J(\nu_1,\nu_2+2,\nu_3-1) - p_1^2\, J(\nu_1,\nu_2+2,\nu_3)\right)
\bigg] \nn \\
&&
\hspace{-0.6cm} 
+ \, \nu_1 \, p_{2 \, \mu} \bigg[
\left(1 +\nu_1 + \nu_3-d/2\right) J(\nu_1+1,\nu_2,\nu_3)
+ (\nu_1+1)\left( J(\nu_1+2,\nu_2,\nu_3-1) - p_2^2\, J(\nu_1+2,\nu_2,\nu_3)\right)
\bigg] \nn \\
&&
\hspace{-0.6cm} 
+ \, \nu_2 \, \bigg[ (\nu_3-1)\,  J_{\mu}(\nu_1,\nu_2+1,\nu_3) + (\nu_2+1) \left( J_{\mu}(\nu_1,\nu_2+2,\nu_3-1)
-  p_1^2\, J_{\mu}(\nu_1,\nu_2 + 2,\nu_3) \right) \bigg] \nn \\
&&
\hspace{-0.6cm} 
- \, \nu_1 \, \bigg[ (\nu_3-1)\,  J_{\mu}(\nu_1+1,\nu_2,\nu_3) + (\nu_1+1) \left( J_{\mu}(\nu_1+2,\nu_2,\nu_3-1) -  p_2^2\, J_{\mu}(\nu_1+2,\nu_2,\nu_3) \right) \bigg] = 0 \, , 
\eea
\normalsize
where the rank-$1$ tensor integral is defined as
\beq
J_{\mu}(\nu_1,\nu_2,\nu_3) = \int \frac{d^d l}{(2\pi)^d}\, 
\frac{l_{\mu}}{(l^2)^{\nu_3}((l+p_1)^2)^{\nu_2}((l-p_2)^2)^{\nu_1}}  = C_1(\nu_1,\nu_2,\nu_3)\, p_{1\,\mu} - C_2(\nu_1,\nu_2,\nu_3)\, p_{2\,\mu}  \, ,
\eeq
with the coefficients given by
\bea
C_1(\nu_1,\nu_2,\nu_3) 
&=& 
\frac{1}{(p_3^2 - p_1^2 - p_2^2)^2 - 4\, p_1^2\, p_2^2 }\, 
\bigg\{ (p_1^2+p_2^2-p_3^2)\, J(\nu_1-1,\nu_2,\nu_3) \nn \\
&& \hspace{-3cm}
- 2\, p_2^2\, J(\nu_1,\nu_2-1,\nu_3)
+
\left( -p_1^2 + p_2^2 + p_3^2 \right)\, J(\nu_1,\nu_2,\nu_3-1)
+ p_2^2\, \left( p_1^2 - p_2^2 + p_3^2 \right)\, J(\nu_1,\nu_2,\nu_3)
\bigg\} \nn \\
C_2(\nu_1,\nu_2,\nu_3)
&=& 
\frac{1}{(p_3^2 - p_1^2 - p_2^2)^2 - 4\, p_1^2\, p_2^2 }\, 
\bigg\{ (p_1^2+p_2^2-p_3^2)\, J(\nu_1,\nu_2-1,\nu_3) \nn \\
&& \hspace{-3cm}
- 2\, p_1^2\, J(\nu_1-1,\nu_2,\nu_3)
+
\left( p_1^2 - p_2^2 + p_3^2 \right)\, J(\nu_1,\nu_2,\nu_3-1)
+ p_1^2\, \left( -p_1^2 + p_2^2 + p_3^2 \right)\, J(\nu_1,\nu_2,\nu_3)
\bigg\}\, . 
\nn \\
\eea
Using the momentum expansion of the tensor integral defined above, we extract from Eq.(\ref{vectorid}) the relation
\small
\bea
&& 
\hspace{-0.7cm}
\nu_2\, (\nu_3-1)\, C_1(\nu_1,\nu_2+1,\nu_3) 
+ \nu_2\,(\nu_2+1)\, \left( C_1(\nu_1,\nu_2+2,\nu_3-1) - p_1^2\, C_1(\nu_1,\nu_2+2,\nu_3) \right) \nn \\
&&
\hspace{-0.7cm}
- \nu_1\, (\nu_3-1)\, C_1(\nu_1+1,\nu_2,\nu_3) 
- \nu_1\,(\nu_1+1)\, \left( C_1(\nu_1+2,\nu_2,\nu_3-1) - p_2^2\, C_1(\nu_1+2,\nu_2,\nu_3) \right) \nn \\
&&
\hspace{-0.7cm}
+\, \nu_2 \bigg[ (\nu_2+1) \left(J(\nu_1,\nu_2+2,\nu_3-1) - p_1^2\, J(\nu_1,\nu_2+2,\nu_3)  \right)
+  \left(1+\nu_2+\nu_3 - d/2 \right)\, J(\nu_1,\nu_2+1,\nu_3) \bigg]= 0 \,, \nn \\
\label{SpecConf}
\eea
\normalsize
together with the corresponding symmetric equation obtained interchanging $(p_1^2, \nu_1) \leftrightarrow (p_2^2, \nu_2)$. \\
This result allows to express integrals in the plane $\nu_1+\nu_2+\nu_3 = \kappa+2$
in terms of those in the two lower ones. In fact, introducing in Eq.(\ref{SpecConf}) and in its symmetric one the explicit expressions for
$C_1$ and $C_2$ we get
\beq \label{ExpandingJ}
J(\nu_1+2,\nu_2,\nu_3) =
\frac{1}{\nu_1\,(\nu_1+1)\,(p_1^2 + p_2^2 - p_3^2)\,p_2^2\,p_3^2}\, \sum_{(a,b,c)} \mathcal C_{(a,b,c)} J(\nu_1+a,\nu_2+b,\nu_3+c)\, ,
\eeq
where the coefficients $\mathcal C_{(a,b,c)}$ are given by 
\bea
&&
\mathcal C_{(0,0,0)} = (\nu_3-1)\, \bigg( (\nu_1+\nu_2)\, p_1^2 - \nu_2\, p_3^2 \bigg)\, , 
\nn \\
&&\mathcal C_{(1,-1,0)} = \nu_1\, (\nu_3-1)\, (p_3^2-p_1^2) \, ,
\nn \\
&&
\mathcal C_{(-1,1,0)} = - \nu_2\, (\nu_3-1)\, p_1^2\, ,\nn
\eea
\bea
&&
\mathcal C_{(0,1,-1)} = \nu_2\, \bigg[ (\nu_2+1)\,p_1^2 - (2+\nu_2-\nu_3)\, p_3^2  \bigg]\, , 
\nn \\
&&
\mathcal C_{(1,0,-1)} = \nu_1 \, (p_1^2 (\nu_1 +1 ) - p_3^2 (\nu_3 -1)) \,,
\nn 
\eea
\bea
&&
\mathcal C_{(2,-1,-1)} = - \nu_1\,(\nu_1+1)\, (p_1^2 - p_3^2)\, , 
\nn \\
&&
\mathcal C_{(-1,2,-1)} = - \nu_2\, (\nu_2+1)\, p_1^2\, , \nn
\eea
\bea%
&&
\mathcal C_{(2,0,-2)} = - \nu_1\, (\nu_1+1)\, p_3^2\, , 
\nn \\
&&
\mathcal C_{(0,2,-2)} = \nu_2\, (\nu_2+1)\, p_3^2\, , 
\nn 
\eea
\bea
&&
\mathcal C_{(1,0,0)} = \nu_1\, \bigg[(p_1^2)^2\, \bigg(\frac{d}{2}-\nu_1-2 \bigg) - p_3^2\, (p_2^2-p_3^2)
\bigg(\frac{d}{2}-\nu_1-\nu_3-1\bigg)
\nn \\
&& \hspace{20mm}
+\, p_1^2\, \bigg( \bigg(1-\frac{d}{2}\bigg)\, p_2^2 +  p_3^2\, (2\, \nu_1 + \nu_3 + 3 - d)\bigg)\bigg]
\nn \\
&&
\mathcal C_{(0,1,0)} = \nu_2\, p_1^2\, \bigg[ (1-\frac{d}{2})\, \left(p_1^2-p_3^2\right) + (\frac{d}{2}-2-\nu_2)\, p_2^2 \bigg]\, ,
\nn \\
&&
\mathcal C_{(0,2,-1)} = - \nu_2\,(\nu_2+1)\, p_1^2\, p_3^2\, , 
\nn \\
&&
\mathcal C_{(-1,2,0)} = \nu_2\,(\nu_2+1)\, (p_1^2)^2\, ,
\nn \\
&&
\mathcal C_{(2,0,-1)} = \nu_1\,(\nu_1+1)\, p_3^2\, (p_1^2 + 2\, p_2^2 - p_3^2)\, , 
\nn \\
&&
\mathcal C_{(2,-1,0)} = \nu_1\, (\nu_1+1)\, p_2^2\, (p_1^2-p_3^2)\, .
\eea
Analogous results hold for $J(\nu_1,\nu_2+2,\nu_3)$ and $J(\nu_1,\nu_2,\nu_3+2)$ if we just make the usual exchanges
$(p_1^2,\nu_1) \leftrightarrow (p_2^2,\nu_2)$ and $(p_1^2,\nu_1) \leftrightarrow (p_3^2,\nu_3)$
both in the integrals and in the coefficients $\mathcal C_{(a,b,c)}$.

%%%%%%%%%%%%%%%%%%%%%%%%%%%%%%%%%%%%%%%%%%%%%%%%%%%%%%%%%%%%%%%%%%%%%%%%%%%%%%%%%%%%%%
\section{Conclusions}

We have shown that the solution of the conformal constraints for a scalar three-point function can be obtained directly in momentum space by solving the differential equations 
following from them. This has been possible having shown that these constraints take the form of a system of two PDE's of generalized hypergeometric type. 
The solution is expressed as a linear combination of four independent Appell's functions. 
The use of the momentum symmetries of the correlator allows to leave free a single multiplicative integration constant to 
parameterize the general solution for any conformal field theory. If this solution is compared with the position space counterpart and its Fourier representation, which is given by a family of 
Feynman master integrals, we obtain the explicit expression of the same integrals in terms of special functions. Our solution coincides with the one found 
by Boos and Davydychev using Mellin-Barnes techniques, which in our case are completely bypassed. 
Having established this link, we have shown that by applying special conformal constraints on the master integral 
representation one obtains new recursion relations. \\
The momentum space approach discussed in this work can be used to treat more complicated correlators. For instance, this method can be employed in the analysis
of three-point functions involving the vector and the energy momentum tensor operators, like $VVV$, $TOO$, $TVV$ and $TTT$, as well as higher order ones,
such as the scalar four-point function, whose general structure has been known for a long time \cite{Polyakov:1970}.
Nevertheless, such a treatment is much more complicated, in the former case due to the tensor nature of the correlators, 
which implies a much more involved set of constraints, in the latter because of the increasing number of independent variables in the partial differential equations. This is left for future work.

\appendix
\section{Appendix. Invariance under the momenta parameterization}
\label{appendix1}

In this appendix we show, exploiting the invariance under rotations and dilatations, that the second of Eq.(\ref{ConformalEqMom}) is independent of which momentum is eliminated. This implies that the two different parameterizations of the special conformal constraints
\bea
&& \sum_{r=1}^{n-1} \left( p_{r \, \mu} \, \frac{\partial^2}{\partial p_{r}^{\nu} \partial p_{r \, \nu}}  - 2 \, p_{r \, \nu} \, 
\frac{\partial^2}{ \partial p_{r}^{\mu} \partial p_{r \, \nu} }    + 2 (\eta_r - d) \frac{\partial}{\partial p_{r}^{\mu}}  + 2 
(\Sigma_{\mu\nu}^{(r)})^{i_r}_{j_r} \frac{\partial}{\partial p_{r \, \nu}} \right) \nn \\
&& \hspace{7cm} \, \times \langle \mathcal O^{i_1}_1(p_1) \ldots  \mathcal O^{j_r}_r(p_r) \ldots \mathcal O^{i_n}_n(p_n) \rangle = 0 \,, \label{specialconf1} \\
&& \sum_{\stackrel{r=1}{r \neq k}}^{n} \left( p_{r \, \mu} \, \frac{\partial^2}{\partial p_{r}^{\nu} \partial p_{r \, \nu}}  - 2 \, p_{r \, \nu} \, 
\frac{\partial^2}{ \partial p_{r}^{\mu} \partial p_{r \, \nu} }    + 2 (\eta_r - d) \frac{\partial}{\partial p_{r}^{\mu}}  + 2 
(\Sigma_{\mu\nu}^{(r)})^{i_r}_{j_r} \frac{\partial}{\partial p_{r \, \nu}} \right) \nn \\
&& \hspace{7cm} \, \times \langle \mathcal O^{i_1}_1(p_1) \ldots  \mathcal O^{j_r}_r(p_r) \ldots \mathcal O^{i_n}_n(p_n) \rangle = 0 \,, \label{specialconf2}
\eea
in which we have respectively removed the dependence on $p_n$ and $p_k$ in terms of the other momenta,
are indeed equivalent. \\
In order to simplify the presentation of the proof we introduce some convenient notations. We define
\bea
G^{i_1 \ldots i_n} &\equiv& \langle \mathcal O^{i_1}_1(p_1) \ldots  \mathcal O^{i_n}_n(p_n) \rangle \,, \nn \\
\mathcal R_{\mu\nu}(p_r) &\equiv& p_{r \, \nu} \frac{\partial}{\partial p_{r}^{\mu}} - p_{r \, \mu} \frac{\partial}{\partial p_{r}^{\nu}} \,, \nn \\
\mathcal D(p_{r}) &\equiv& - p_{r \, \nu} \frac{\partial}{\partial p_{r \, \nu}} - d \,, \nn \\
\mathcal K_{\mu}(p_{r}, \eta) &\equiv& p_{r \, \mu} \, \frac{\partial^2}{\partial p_{r}^{\nu} \partial p_{r \, \nu}}  - 2 \, p_{r \, \nu} \, 
\frac{\partial^2}{ \partial p_{r}^{\mu} \partial p_{r \, \nu} }    + 2 (\eta - d) \frac{\partial}{\partial p_{r}^{\mu}} \,
\eea
and preliminarily derive two constraints, which will be used in the following, emerging from the invariance under rotations and scale transformations respectively. \\
Using the same procedure described in section 2, we find the constraint coming from rotational invariance
\bea
\sum_{r = 1}^{n-1} \mathcal R_{\mu\nu}(p_r) G^{i_1 \ldots i_n} - \sum_{r = 1}^{n}( \Sigma_{\mu\nu}^{(r)})^{i_r}_{j_r} G^{i_1 \ldots j_r \ldots i_n} = 0 \,,
\eea
from which, differentiating with respect to $p_{k \, \nu}$, we obtain
\bea
\label{diffRot}
\bigg[ \sum_{r = 1 \,, r \neq k}^{n-1} \mathcal R_{\mu\nu}(p_r) \frac{\partial}{\partial p_{k \, \nu}}  +   \mathcal F_{\mu}(p_k) \bigg] G^{i_1 \ldots i_n} 
- \sum_{r = 1}^{n}( \Sigma_{\mu\nu}^{(r)})^{i_r}_{j_r} \frac{\partial}{\partial p_{k \, \nu}} G^{i_1 \ldots j_r \ldots i_n} = 0 \,, 
\eea
where $\mathcal F_{\mu}(p_k)$ is defined by
\bea
\label{tensorF}
\mathcal F_{\mu}(p_k) = (d-1) \frac{\partial}{\partial p_{k}^{\mu}} + p_{k}^\nu  \frac{\partial^2}{\partial p_{k}^{\nu} \, \partial p_{k}^{\mu}} - p_{k \, \mu}  \frac{\partial^2}{\partial p_{k}^{\nu} \, \partial p_{k \, \nu}} \,.
\eea
Another useful relation can be obtained differentiating the dilatation constraint in the first of Eq.(\ref{ConformalEqMom}) with respect to $p_{k\, \mu}$
\bea
\label{diffDil}
\bigg[ \sum_{r = 1\,, r \neq k}^{n-1} \left( \mathcal D(p_r) + \eta_r \right) \frac{\partial}{\partial p^\mu_k}  + (\eta_k + \eta_n - d- 1) \frac{\partial}{\partial p^\mu_k} - p_{k}^\nu \frac{\partial^2}{\partial p^\mu_k \partial p^\nu_k} \bigg] G^{i_1 \ldots  i_n} = 0 \,.
\eea
Having derived all the necessary relations, we can proceed with the proof of equivalence between Eqs.(\ref{specialconf1}) and (\ref{specialconf2}). We remove from Eq.(\ref{specialconf1}) the $k-th$ term containing the spin matrix $(\Sigma_{\mu\nu}^{(k)})^{i_k}_{j_k}$ using Eq.(\ref{diffRot}) 
\bea
\label{specialconftemp1}
&& \bigg[ \sum_{r = 1}^{n-1} \mathcal K_\mu(p_r, \eta_r) + 2 \sum_{r=1 \,, r \neq k}^{n-1} \mathcal R_{\mu\nu}(p_r) \frac{\partial}{\partial p_{k \, \nu}} + 2 \mathcal F_\mu(p_k) \bigg] G^{i_1 \ldots  i_n} \nn \\
&& + \,  2 \sum_{r=1 \,, r \neq k}^{n-1}( \Sigma_{\mu\nu}^{(r)})^{i_r}_{j_r} \left( \frac{\partial}{\partial p_{r \, \nu}} - \frac{\partial}{\partial p_{k \, \nu}}\right) G^{i_1 \ldots j_r \ldots i_n} 
- 2( \Sigma_{\mu\nu}^{(n)})^{i_n}_{j_n} \frac{\partial}{\partial p_{k \, \nu}} G^{i_1 \ldots  j_n}  = 0 \,,
\eea
and then we combine the $k-th$ operator $\mathcal K_\mu(p_k, \eta_k)$ with the $\mathcal F_\mu(p_k)$ contribution as
\bea
\mathcal K_{\mu}(p_k, \eta_k) + 2 \mathcal F_\mu(p_k) = - p_{k \, \mu} \frac{\partial^2}{\partial p_{k \, \nu} \partial p_k^\nu} + 2 (\eta_k - 1) \frac{\partial}{\partial p_k^\mu} \,.
\eea
Using Eq.(\ref{diffDil}) we rewrite the previous equation as
\bea
\mathcal K_{\mu}(p_k, \eta_k) + 2 \mathcal F_\mu(p_k) = - \mathcal K_\mu(p_k, \eta_n) - 2 \sum_{r = 1\,, r \neq k}^{n-1} \left( \mathcal D(p_r) + \eta_r \right) \frac{\partial}{\partial p_{k}^\mu}
\eea
so that Eq.(\ref{specialconftemp1}) can be recast in the following form
\bea
\label{specialconftemp2}
&& \bigg\{ \sum_{r=1 \,, r \neq k}^{n-1} \bigg[ \mathcal K_\mu(p_r , \eta_r) + 2 \, \mathcal R_{\mu\nu}(p_r) \frac{\partial}{\partial p_{k \, \nu}} - 2 \left( \mathcal D(p_r) + \eta_r \right) \frac{\partial}{\partial p_{k \, \mu}}  \bigg] - \mathcal K_\mu(p_k, \eta_n) \bigg\} G^{i_1 \ldots  i_n}   \nn \\
&&
+ \,  2 \sum_{r=1 \,, r \neq k}^{n-1}( \Sigma_{\mu\nu}^{(r)})^{i_r}_{j_r} \left( \frac{\partial}{\partial p_{r \, \nu}} - \frac{\partial}{\partial p_{k \, \nu}}\right) G^{i_1 \ldots j_r \ldots i_n} 
- 2( \Sigma_{\mu\nu}^{(n)})^{i_n}_{j_n} \frac{\partial}{\partial p_{k \, \nu}} G^{i_1 \ldots  j_n}  = 0 \,.
\eea
In order to show the equivalence of Eq.(\ref{specialconftemp2}) with Eq.(\ref{specialconf2}) it is necessary to perform a change of variables from the independent set of momenta $(p_1, \ldots p_{n-1})$ to the new independent one $(p_1 \ldots p_{k-1}, p_{k+1}, \ldots p_n)$ from which $p_k$ has been removed using momentum conservation $p_k = - \sum_{r=1 \,, r \neq k}^ n p_r$. In this respect all the derivatives appearing in Eq.(\ref{specialconftemp2}) must be replaced according to
\bea
&& \frac{\partial}{\partial p_{r \, \mu}} \rightarrow \frac{\partial}{\partial p_{r \, \mu}} - \frac{\partial}{\partial p_{n \, \mu}}  \qquad \qquad \mbox{for} \quad r = 1, \ldots n-1 \quad \mbox{with} \quad r \neq k \,, \nn \\
&& \frac{\partial}{\partial p_{k \, \mu}} \rightarrow - \frac{\partial}{\partial p_{n \, \mu}} \qquad \qquad \mbox{for} \quad r = k \,.
\eea
It is just matter of tedious algebraic manipulations to show that the operators in curly brackets in Eq.(\ref{specialconftemp2}) simplify, after the change of variables, to give $\sum_{r = 1 \,, r \neq k}^n \mathcal K_\mu(p_r, \eta_r)$, while the two spin matrices sum up together and we are left with
\bea
\sum_{r = 1 \,, r \neq k}^n \mathcal K_\mu(p_r, \eta_r) G^{i_1, \ldots i_n} + 2 \sum_{r = 1 \,, r \neq k}^n( \Sigma_{\mu\nu}^{(r)})^{i_r}_{j_r} \frac{\partial}{\partial p_{r \, \nu}} G^{i_1 \ldots j_r \ldots i_n} = 0
\eea
which is exactly Eq.(\ref{specialconf2}), where, now, $G^{i_1, \ldots i_n}$ is understood to be a function of the independent momenta $(p_1 \ldots p_{k-1}, p_{k+1}, \ldots p_n)$. This completes our derivation proving the independence of the special conformal constraints on which momentum is removed using the momentum conservation equation.

\section{Conformal constraints on two-point functions}
\label{AppTwoPoint}
In this appendix we provide some details on the solutions of the conformal constraints for the two-point functions with conserved vector and tensor operators. \\
In the first case the tensor structure of the two-point function is uniquely fixed by the transversality condition $\partial^\mu V_\mu$ as
\bea
G_V^{\alpha \beta}(p) = f(p^2) t^{\alpha \beta}(p)\,, \qquad \mbox{with} \quad t^{\alpha\beta}(p) = p^2 \eta^{\alpha\beta} - p^{\alpha} p^{\beta} \,.
\eea 
For the sake of simplicity, we have employed in the previous equation a slightly different notation with respect to Eq.(\ref{TwoPointVector0}), which, anyway, can be recovered with the identification $f(p^2) = f_V(p^2)/p^2$. \\
In order to exploit the invariance under scale and special conformal transformations it is useful to compute first and second order derivatives of the $t^{\alpha \beta}$ tensor structure. In particular we have
\bea
\label{TDerivatives}
t_1^{\alpha \beta, \mu}(p) &\equiv& \frac{\partial }{\partial p_\mu} t^{\alpha \beta}(p) = 2 \, p^{\mu} \eta^{\alpha \beta} - p^{\alpha} \eta^{\mu \beta} - p^{\beta} \eta^{\mu \alpha} \,, \nn \\
t_2^{\alpha \beta, \mu \nu}(p) &\equiv& \frac{\partial^2 }{\partial p_\mu \, \partial p_\nu} t^{\alpha \beta}(p) = 2 \, \eta^{\mu \nu} \eta^{\alpha \beta} - \eta^{\nu \alpha} \eta^{\mu \beta} - \eta^{\nu \beta} \eta^{\mu \alpha} \,, 
\eea
with the properties
\bea
&& p_\mu t_1^{\alpha \beta, \mu}(p) = 2 \, t^{\alpha \beta}(p) \,, \qquad     t_1^{\alpha \beta, \alpha}(p) = - (d - 1) p^\beta \,, \nn \\
&& p_\mu t_2^{\alpha \beta, \mu \nu}(p) = t_1^{\alpha \beta, \nu}(p) \,, \qquad    t_2^{\alpha \beta, \mu \mu}(p) = 2(d-1) \eta^{\alpha \beta} \,. 
\eea
As we have already mentioned, the invariance under scale transformations implies
\bea
\label{FandLambda}
f(p^2) = (p^2)^\lambda \qquad \qquad \mbox{with} \quad \lambda = \frac{\eta_1 + \eta_2 - d}{2} -1 \,,
\eea
which can be easily derived from the first order differential equation in (\ref{ConformalEqMomTwoPoint}) using Eq.(\ref{TDerivatives}). Having determined the structure of the scalar function $f(p^2)$, one can compute the derivatives appearing in the second of Eq.(\ref{ConformalEqMomTwoPoint}), namely the constraint following from the invariance under the special conformal transformations
\bea
\label{GDerivatives}
\frac{\partial}{\partial p_\mu} G_V^{\alpha \beta}(p) &=& (p^2)^{\lambda - 1} \bigg[ 2 \lambda \, p^\mu t^{\alpha \beta}(p) + p^2 \, t_1^{\alpha\beta, \mu}(p) \bigg] \,, \nn \\
\frac{\partial^2}{\partial p_\mu \, \partial p_\nu} G_V^{\alpha \beta}(p) &=&(p^2)^{\lambda - 2} \bigg[ 4 \lambda (\lambda -1) p^{\mu} p^{\nu} t^{\alpha \beta}(p)  + 2 \lambda p^2 \eta^{\mu\nu}   t^{\alpha \beta}(p) + 2 \lambda  p^2 p^{\mu} t_1^{\alpha\beta,\nu}(p)  \nn \\
&& \qquad +  \, 2 \lambda  p^2 p^{\nu} t_1^{\alpha\beta,\mu}(p) + (p^2)^2 t_2^{\alpha\beta, \mu\nu}(p) 
\bigg] \,,
\eea
where the definitions in Eq.(\ref{TDerivatives}) have been used.
Concerning the spin dependent part in Eq.(\ref{ConformalEqMomTwoPoint}), we use the spin matrix for the vector field, which, in our conventions, reads as
\bea
( \Sigma_{\mu\nu}^{(V)})^{\alpha}_{\beta} = \delta_{\mu}^{\alpha} \, \eta_{\nu \beta} - \delta_{\nu}^{\alpha} \, \eta_{\mu \beta} \,,
\eea
and obtain
\bea
\label{SigmaPart}
2( \Sigma_{\mu\nu}^{(V)})^{\alpha}_{\rho} \frac{\partial}{\partial p_\nu} G_V^{\rho\beta}(p) = - (p^2)^{\lambda- 1} \bigg[ 2 \lambda \, p^\alpha {t_{\mu}}^{\beta}(p) + (d-1)p^2 p^\beta \delta_\mu^\alpha  + p^2 {t_{1 \, \mu}}^{\beta, \alpha}(p) \bigg] \,.
\eea
Employing the results derived in Eq.(\ref{GDerivatives}) and Eq.(\ref{SigmaPart}), we have fully determined the special conformal constraint on the two-point vector function.
Then we can project the second of Eq.(\ref{ConformalEqMomTwoPoint}) onto the three independent tensor structures $p_\mu \eta_{\alpha \beta}, p_\alpha \eta_{\mu \beta}, p_\beta \eta_{\alpha \mu}$, and setting $\lambda$ to the value given in Eq.(\ref{FandLambda}), we finally obtain three equations for the scale dimensions $\eta_i$ of the vector operators
\bea
\begin{cases}
(\eta_1 - \eta_2) (\eta_1 + \eta_2 - d) = 0 \,, \nn \\
\eta_1 - d +1 = 0 \,, \nn \\
\eta_2 - d +1 = 0 \,. \nn
\end{cases}
\\
\eea
The previous system of equations can be consistently solved only for $\eta_1 = \eta_2 = d -1$, as expected. This completes our derivation of the vector two-point function which, up to an arbitrary multiplicative constant, can be written as in Eq.(\ref{TwoPointVector}).

The characterization of the two-point function with a symmetric, traceless and conserved rank-2 tensor follows the same lines of reasoning already explained in the vector case. These conditions (see Eq.(\ref{EMTconditions})) fix completely the tensor structure of the two-point function as
\bea
G_T^{\alpha\beta\mu\nu}(p) = g(p^2) \, T^{\alpha\beta\mu\nu}(p) 
\eea
with
\bea
T^{\alpha\beta\mu\nu}(p) =  \frac{1}{2} \bigg[ t^{\alpha\mu}(p) t^{\beta\nu}(p) + t^{\alpha\nu}(p) t^{\beta\mu}(p) 
\bigg] 
- \frac{1}{d-1} t^{\alpha\beta}(p) t^{\mu\nu}(p) \,.
\eea
In order to recover the convention used in section \ref{TwoPointSection}, notice that $g(p^2) = f_T(p^2)/(p^2)^2$. \\
As in the previous case, we give the first and second order derivatives of the $T^{\alpha\beta\mu\nu}(p)$ tensor structure
\bea
\label{TTDerivatives}
T_1^{\alpha\beta\mu\nu, \rho}(p) &\equiv& \frac{\partial}{\partial p_\rho} T^{\alpha\beta\mu\nu}(p) = \frac{1}{2} \bigg[ t_1^{\alpha\mu, \rho}(p) t^{\beta\nu}(p) 
+ t^{\alpha\mu}(p) t_1^{\beta\nu, \rho}(p) + \left( \mu \leftrightarrow \nu \right) \bigg] \nn \\
&& - \frac{1}{d-1} \bigg[ t_1^{\alpha\beta, \rho}(p) t^{\mu\nu}(p) + t^{\alpha\beta}(p) t_1^{\mu\nu, \rho}(p) \bigg] \,, \nn \\
T_2^{\alpha\beta\mu\nu, \rho\sigma}(p) &\equiv& \frac{\partial}{\partial p_\rho \, \partial p_\sigma} T^{\alpha\beta\mu\nu}(p) = \frac{1}{2} \bigg[ 
t_2^{\alpha\mu, \rho \sigma}(p) t^{\beta\nu}(p) + t_1^{\alpha\mu, \rho}(p) t_1^{\beta\nu, \sigma}(p) + t_1^{\alpha\mu, \sigma}(p) t_1^{\beta\nu, \rho}(p) \nn \\
&& +  \, t^{\alpha\mu}(p) t_2^{\beta\nu, \rho \sigma}(p)+ \left( \mu \leftrightarrow \nu \right) \bigg]  \nn \\
&& - \frac{1}{d-1} \bigg[  t_2^{\alpha\beta, \rho\sigma}(p) t^{\mu\nu}(p) +   t_1^{\alpha\beta, \rho}(p) t_1^{\mu\nu, \sigma}(p) + (\mu\nu) \leftrightarrow (\alpha\beta)  \bigg] \,,
\eea
together with some of their properties
\bea
p_\rho T_1^{\alpha\beta\mu\nu, \rho}(p) = 4 \, T^{\alpha\beta\mu\nu}(p) \,, \qquad
p_\rho T_2^{\alpha\beta\mu\nu, \rho \sigma}(p) = 3 \, T_1^{\alpha\beta\mu\nu, \sigma}(p)  \,.
\eea
As we have already stressed, the first of Eq.(\ref{ConformalEqMomTwoPoint}) defines the scaling behavior of the two-point function, providing, therefore, the functional form of $g(p^2)$ which is given by
\bea
g(p^2) = (p^2)^\lambda \qquad \mbox{with} \quad \lambda = \frac{\eta_1 +\eta_2 -d}{2} -2 \,.
\eea
On the other hand, the second of Eq.(\ref{ConformalEqMomTwoPoint}), namely the constraint from the special conformal transformations, fixes the scaling dimensions of the tensor operators. In this case the spin connection is given by
\bea
(\Sigma_{\mu\nu}^{(T)})^{\alpha \beta}_{\rho \sigma } = \left( \delta_{\mu}^{\alpha} \, \eta_{\nu \rho} - \delta_{\nu}^{\alpha} \, \eta_{\mu \rho} \right) \delta_{\sigma}^{\beta}
+ \left( \delta_{\mu}^{\beta} \, \eta_{\nu \sigma} - \delta_{\nu}^{\beta} \, \eta_{\mu \sigma} \right) \delta_{\rho}^{\alpha} \,.
\eea
The algebra is straightforward but rather cumbersome due to the proliferation of indices. Here we give only the final result, which can be obtained projecting Eq.(\ref{ConformalEqMomTwoPoint}), making use of Eq.(\ref{TTDerivatives}), onto the different independent tensor structures
\bea
\begin{cases}
(\eta_1 - \eta_2) (\eta_1 + \eta_2 - d) = 0 \,, \nn \\
\eta_1 - d = 0 \,, \nn \\
\eta_2 - d = 0 \,, \nn
\end{cases}
\\
\eea
which implies the solution $\eta_1 = \eta_2 = d$, as described in Eq.(\ref{TwoPointEmt}).

%%%%%%%%%%%%%%%%%%%%%%%%%%%%%%%%%%%%%%%%%%%%%%%%%%%%%%%%%%%%%%%%%%%%%%%%%%%%%%%%%%%%%%

%%%%%%%%%%%%%%%%%%%%%%%%%%%%%%%%%%%%%%%%%%%%%%%%%%%%%%%%%%%%%%%%%%%%%%

\end{document}